\documentclass[authoryear,preprint,review,12pt]{elsarticle}

\usepackage{graphicx}
\usepackage{amssymb}
\usepackage{makeidx}
\usepackage{bm}
\usepackage{epsfig}
\usepackage{psfrag}
\usepackage{lineno}
\usepackage{setspace}

\usepackage{amssymb}
\usepackage{natbib}
\journal{Physics of the Earth and Planetary Interiors}

\def\dsize{\displaystyle}
\def\E{\mathop{\rm E}\nolimits}
\def\Ro{\mathop{\rm Ro}\nolimits}
\def\Rm{\mathop{\rm R_m}\nolimits}
\def\Re{\mathop{\rm Re}\nolimits}
\def\Ra{\mathop{\rm Ra}\nolimits}

\def\Pr{\mathop{\rm Pr}\nolimits}

\begin{document}

\begin{frontmatter}



\title{Effects of anisotropy in geostrophic turbulence}


\author[ph]{P.~Hejda}
\ead{ph@ig.cas.cz}
\author[mr]{M.~Reshetnyak\corref{cor1}}
\ead{m.reshetnyak@gmail.com}

\address[ph]{Institute of Geophysics, Academy of Sciences, 141 31
Prague, Czech Republic}

\address[mr]{
Institute of the Physics of the Earth, Russian Acad.~Sci,
 123995 Moscow, Russia}

\vskip 1cm \cortext[cor1]{ Corresponding author}

\begin{abstract}
The Boussinesq model of convection in a flat layer  with heating
from below is considered. We analyze the effects of anisotropy
caused by rapid rotation in physical and wave spaces and
demonstrate  the suppression of energy transfer by rotation. We
also examine the
 structure of the wave triangle in nonlinear interaction. The
 range of parameters
 is adapted to the models of convection in the  geodynamo.
\end{abstract}


\end{frontmatter}



\section{Introduction}
\label{s1} Approximation in the form of homogeneous and isotropic
turbulence is quite crude for describing many geophysical
applications.
 Usually, there are two reasons why this approximation can be
 violated. One of them is a rapid daily rotation characterized by a
 small Rossby number, $\Ro\ll 1$.
 This case is typical for problems of meteorology and physics of the
 ocean where its influence is principal \citep{Gill}.
 However, the most critical regime with an extremely small Rossby number
 occurs in the liquid planetary cores.
 There, the Coriolis force is three orders of magnitude larger than the
 nonlinear term, which  changes the balance of the forces in the
 Navier-Stokes equation
 and the spectral properties of turbulence  for large Reynolds
 numbers, $\Re$.
 Another source of anisotropy of convection in the liquid core is
 a strong magnetic field which, in combination with
 rotation, leads to the plate-like structure of convective patterns
 \citep{BragMeytl,Matsushima,Donald}.  In the present paper  we
 consider the former
  source of anisotropy related to rapid rotation.

The introduction of rotation leads to substantial rearranging of
the flow, both in the physical and in  the wave spaces.  In spite
of this, the Coriolis force itself does not produce work.  It can
redistribute the energy between the scales and lead to inverse
cascades, which is known in direct numerical simulations (DNS )
as an increase of the kinetic energy on large scales
\citep{Hossain} predicted in the renormalization group theory
\citep{McComb}. As a result,
 the slope of the spectrum of the kinetic energy changes from $-5/3$
 to  $-2$ \citep{Zhou5,Constantin}. This change is closely related to
 the break of the energy transfer  over the spectrum  \citep{Zhou5}.

Let us proceed to the thermal convection problem, when an
additional equation of  heat transfer  leads to the
transformation  of cellular convection to the cyclonic
convection. This problem has been studied thoroughly in the
pioneering works on thermal convection
\citep{Ch,Roberts65,Busse70}, as well as in later papers on the
geodynamo, see  \citep{HolRud} for references.

It appears that rapid rotation causes a new kind of balance
between pressure and the Coriolis force (geostrophic balance)
\citep{Pedlosky}: $1_z\times{\bf V}\sim \nabla p $, which leads
to small  gradients along the axis of rotation $\bf z$: $\dsize
{\partial {\bf V}\over \partial z}\sim 0$~\footnote{Note that
balance of the potential part of the Coriolis force and the
pressure holds.}.  As a result, cyclones and anticyclones
prolonged  along $\bf z$ of a small diameter appear.  If one
takes the parameters of the Earth's liquid core and decreases the
amplitude of  the
 heating sources  to the onset of convection, the diameter of
 cyclones will then be $d_c\sim \E^{1/3}{\rm L_z}\approx 10^{-5}{\rm
 L_z}$, where $\E\sim 10^{-15}$ is the  Ekman number and ${\rm L_z}$
 is the height of the cyclone. In reality, the heat sources
 are quite larger, and there is a wave
 packet of such cyclones.  However, estimates of the order of the energy
 indicates that, at least for the first 3-4 orders of the wave number $k$,
 geostrophic balance takes place. This exceeds the extent of geomagnetic
 spectra and should be taken into account in geodynamo models.
 This was already started in the development of the   full tensor
 approach to anisotropic viscosity \citep{Phillips}, numerical
 simulations  \citep{Matsushima} and estimates of the anisotropy of
 heat mass transfer
\citep{Donald,Matsui}.

The other role of the  Coriolis force, which will be the subject
of our paper, is to control the  energy transfer over the
spectrum in the system.  Whereas if before the nonlinear term was
of the same order with the pressure, which could not block the
curl part of the term, then with the Coriolis force the  pressure
and other potential forces (e.g., the Archemedean force) block
the potential part of the nonlinear term. And the rest of the
curl parts of the forces and Coriolis force block the nonlinear
term. As a result, the spectrum becomes steeper.  That is what we
have on small scales, $l \ll d_c$. At the same time,  on scales
 $l \gg d_c$,
 the system can be in a state of statistical equilibrium: dissipation
 is negligible, and over the long term the energy exchange  between
 the scales     is small. It appears that the estimate of  only the spectra
 of kinetic energy is not enough to resolve the states with and without
 rotation and intensive heat sources, and one needs to keep track of the fluxes
 of kinetic energy over the spectrum, which is considered below.

Hereinafter, on the example of an 3D model of thermal convection
in a plane layer, we consider the properties of anisotropy caused
by rotation both in the physical and especially in the wave spaces
started in \citep{HR08} typical for geodynamo regimes and compare
these results with nonrotating convection.

\section{The Boussinesq model}\label{BM}
\subsection{Equations in physical space}
The thermal convection equations for an incompressible fluid
  ($\nabla\cdot{\bf V}=0$) in a layer of   height  $\rm L$
 rotating with angular velocity $\Omega$  in a Cartesian system of
  coordinates $(x,\,y,\,z)$
 in its traditional dimensionless form can
be expressed as follows:
\begin{equation}
\begin{array}{l}\dsize
      \E\Pr^{-1}\left[\frac {{\partial} {\bf V}} {\partial t}+ \left({\bf
V}\cdot \nabla\right) {\bf V}\right] = -\nabla { P}  -{\bf
{1}_z}\times{\bf V} + \\ \qquad\qquad \Ra { T} \,z{\bf{1}_z}+
 \E\Delta {\bf V}
\\ \dsize
\frac{\partial { T}} {\partial t}+\left({\bf V}\cdot\nabla\right)
\left({ T}+{ T}_0\right)= \Delta{T}.
\end{array}\label{sys0}
\end{equation}
Velocity $\bf V$, pressure $P$ and the typical diffusion time $t$
are measured in units of $ \rm \kappa/L$,    $\rm
\rho\kappa^2/L^2$ and $\rm L^2/\kappa$, respectively, where
$\kappa$ is the thermal  diffusivity, $\rho$ is the density,
$\dsize \Pr=\frac{\kappa}{ \nu}$ is the Prandtl number,
  $ \dsize \E =\rm  \frac{\nu}{ 2\Omega L^2}$ is the
Ekman number, and $\nu$ is the kinematic viscosity. $\dsize \Ra\rm
=\frac{\alpha g_0\delta {\it T} {L}}{  2\Omega\kappa}$ is the
modified Rayleigh number, $\alpha$ is the coefficient of volume
expansion, $\delta T$ is the unit of temperature, for more
details see  \citep{Jones} ,
 $g_0$ is the gravitational acceleration, and $T_0=1-z$ is the heating from below.
 The problem is closed with periodical boundary conditions in the $(x,\, y)$ plane. In the $z$-direction, we use
  simplified conditions  \citep{Cattaneo}:
$T=0$,
  $\dsize V_z={\partial V_x\over \partial z}=
 {\partial V_y\over \partial z}=0$   at $z=0,\,1$.

\subsection{Equations in wave space}
To solve problem (\ref{sys0})
 we apply the pseudo-spectral approach
\citep{Or}. We follow the modification of the original  approach 
that has been frequently used in thermal convection and
geodynamo: we change FFT in z-direction to the sine-/cosine-decomposition
leaving the
full FFT in the horizontal directions. This let us to
provide the mentioned above boundary conditions in z-direction ,
see also,  \citep{JR, Buffett}.
 The equations are solved in the wave space. To
calculate the non-linear terms one needs to
 apply the inverse Fourier transform, then calculate the product in physical space,
   apply the Fourier transform of the
  product, and finally calculate the derivatives in wave space.
After eliminating the pressure using the
 divergence-free  condition ${\bf k}\cdot {\bf V}=0$  we arrive at:
\begin{equation}\begin{array}{l}\dsize
\dsize
 \E\left[\Pr^{-1}
 {\partial {\bf V}\over\partial t}
  +k^2{\bf V}\right]_{\bf k}
= {\bf k} {\cal P}_{\bf k}+{\bf F}_{\bf k}
\\ \\ \dsize
\left[ {\partial { T}\over\partial t}+k^2 T\right]_{\bf k}=
-\left[\left({\bf V}\cdot\nabla\right){ T}+V_r\right]_{\bf k}
\end{array}\label{sys_s}
\end{equation}
with
\begin{equation}\begin{array}{l}\dsize

{\cal P}_{\bf k}=-{ {\bf k}\cdot {\bf F}_{\bf k}\over k^2}, \qquad
\dsize k^2=k_\beta k_\beta,\qquad \beta=1\dots 3
\\\\
\dsize {\bf F}_{\bf k}=\Big[ \Pr^{-1} {\bf V} \times
\left(\nabla\times {\bf V} \right)+ \Ra T{\bf{1}_z}- {\bf
1_z}\times {\bf V}
 \Big]_{\bf k}.
\end{array}\label{P123}
\end{equation}
For integration in time we use the explicit
 Adams-Bashforth (AB2) scheme for non-linear terms.
 The linear terms are treated using the Crank-Nicolson (CN) scheme. To resolve
  the diffusion terms, we use the well-known trick of changing the variable which helps to increase
   the time step significantly. Consider equation
\begin{equation}\begin{array}{l}\dsize
{\partial A  \over \partial t}+k^2A=U .
\end{array}\label{tr1}
\end{equation}
Alter it to read
\begin{equation}\begin{array}{l}\dsize
{\partial A e^{k^2\gamma t} \over \partial t}=U\, e^{k^2\gamma t}
\end{array}\label{tr2}
\end{equation}
and then apply the CN scheme.

The most time-consuming part of our MPI code are the Fast Fourier
transforms. To make our code more efficient we use various
modifications of known FFT algorithms, which take into account
special kinds of symmetry of the fields. The optimal number of
processors for the $128^3$  grids is $n\sim 50$. Scalability
tests have demonstrated even the presence of superacceleration if
the number of processors $<n$.

\section{Fluxes in  $k$-space} \label{p2}
Now we introduce some diagnostic tools which are very helpful for
the further analysis. To analyze the energy transfer in the wave
space, we follow \cite{Frisch}. Let us decompose the physical
field
 $f$  into a sum of low-frequency and high-frequency counterparts:
      $f({\bf r})=f^<({\bf r})+f^>({\bf r})$, where
      \begin{equation}\begin{array}{l}\dsize
f^<({\bf r})=\sum\limits_{|k|\le K} \widehat{f}_k\,e^{i{\bf
k}{\bf r}}, \qquad f^>({\bf r})=\sum\limits_{|k|>K}
\widehat{f}_k\,e^{i{\bf k}{\bf r}}.
\end{array}\label{sys1}
\end{equation}
For any periodical  functions $f$ and  $g$ one has the relation
\citep{Frisch}:
  \begin{equation}\begin{array}{c}\dsize
<{\partial f \over\partial x}>=0,\qquad <{\partial g
\over\partial x}>=0, \\ \\  \dsize <g{\partial f \over\partial
x}>= -<f{\partial g \over\partial x}>,\qquad <f^>g^<>=0,
\end{array}\label{sys2}
\end{equation}
where
\begin{equation}\begin{array}{c}\dsize
<f({\bf r})>={\cal V}^{-1}\int\limits_{\cal V} f({\bf r})\, d{\bf
r}^3
\end{array}\label{sys3}
\end{equation}
stands for averaging
 $f$ over volume  ${\cal V}$.
Multiplying the Navier-Stokes equation by ${\bf V}^<$ one derives
the equation for $E^<$ in sphere $k<K$:
       \begin{equation}\begin{array}{l}\dsize
2^{-1}\E\Pr^{-1}\left[ {\partial \Big< { V_i^<
V_i^<}\Big>\over\partial t}+ \Pi(K) \right]
  =
     \Ra \Big< { T}^< \,V_z^<\Big>-
   \E\Big<\left(\nabla {\bm \omega} ^<\right)^2\Big>,
\end{array}\label{sys10a}
\end{equation}
 where the integral flux of kinetic energy from the region  $k>K$ to  $k\le K$  reads
\begin{equation}\begin{array}{l}\dsize
\Pi(K)=\Big<  V_i^<\cdot \Big(  V_j\cdot \nabla_j\Big) V_i\Big>,
\end{array}\label{sys100a}
\end{equation}
  with a summation over the repeating indexes $i=1\dots 3$. Note, that flux of the Coriolis force is zero: $-({\bf V}\times {\bf V^<})_z=
-({\bf V^<}\times {\bf V^<})_z -({\bf V^>}\times {\bf V^<})_z=0
$. The first term in sum is zero due to vector identity and the
second due to (5).
  Introducing the local flux  $T_K$:
       \begin{equation}\begin{array}{l}\dsize
     T_K(k)= -{\partial \Pi(k) \over\partial k},\qquad
     \int\limits_{k=0}^{\infty}  T_K(k)\, dk=0,
\end{array}\label{sk1}
\end{equation}
where we have changed K to k, leads to the obvious relation for
$E$ in $k$-space:
\begin{equation}\begin{array}{l}\dsize
 {\partial E(k)\over\partial t}
= {   T}_K(k)+F(k)+D(k),
\end{array}\label{s1}
\end{equation}
where $\dsize E(k)={1\over 2}{\partial \over \partial k }<{\bf
V}_k^2>$ is the change of the kinetic energy
 at  $ k$,
$\dsize F(k)= {\Ra\Pr\over\E}  {\partial \over \partial k }<T
V_{z}^{<} >$  is the work produced by the Archemedean force, and
$D(k)=-\Pr k^2 E(k)$ is the  viscous dissipation. Equation
  (\ref{s1}) оdescribes the current flux of the kinetic energy through
 wave number  $k$.

To describe a triad interaction mechanism, we can also ask the
question: ``What is the form of the energy balance equation which
describes
 the transfer of the energy from wave numbers $Q$ and $P$ to $K$?''
 A similar manipulation with the Navier-Stokes equation yields:
 \begin{equation}\begin{array}{l}\dsize
 {\partial E(K)\over\partial t}
= {   T}_3+A(K)+D(K),
\end{array}\label{s2}
\end{equation}
 where
$E(K)$, $D(K)$ have the same form as in  $(\ref{s1})$, and
$\dsize {   T}_3=\Big<  V_i(K)\cdot \Big(  V_j(P)\cdot
\nabla_j\Big) V_i(Q)\Big>$, $\dsize A(K)={\Ra\Pr\over\E}
<T(K)V_z(K) >$.

It is also useful to introduce function   $\dsize {
T}_2(K,Q)=\int T_3(K,Q,P) \,d P$. In general,  one needs an
explanation. It is possible to show
 that  $T_2$  is the energy flux  from harmonic  $Q$  to harmonic  $K$ (see the references in \citep{ALEX07}).  The analysis of
  $\dsize {   T}_{2}$ allows us to estimate whether the flux is local or not, but not the locality of the interaction itself. The study of  $T_3$
 allow us to recover the full structure of the wave triangle and to explain whether the interaction itself is local.
 We also note some useful properties of function  $T_2$:
 for the arbitrary periodical (or random homogeneous) nondivergent  fields   ${\bf u}(Q)$,  ${\bf w}(K) $and  $\bf  V $ one has
 \citep{Verma,ALEX05}:
$ {T_2^{u w}}(Q,\,K)=-{T_2^{w u}}(K,\, Q) $, where $ {T_2^{u
w}}(Q,\,K)=\Big<  u_i(K)\cdot \Big(  V_j\cdot \nabla_j\Big)
w_i(Q)\Big>$,   $ T_{ w u}(Q,\,K)=\Big<  w_i(Q)\cdot \Big(
V_j\cdot \nabla_j\Big) u_i(K)\Big>$, which corresponds to
  the balance of the energy received by shell  $K$  from shell  $Q$ to that one given by Q  to $K$.
 In the next paragraph, we consider the properties of fluxes  $ {   T}_K $, ${   T}_{2}  $, $T_3$ on an example of
 model  $(\ref{sys0})$ and find how they change when rotation is switched on.

\section{Basic properties of the fields}
\begin{figure}[th!]
\vskip -43.0cm
\begin{minipage}[t]{.35\linewidth}
\hskip -2cm \epsfig{figure=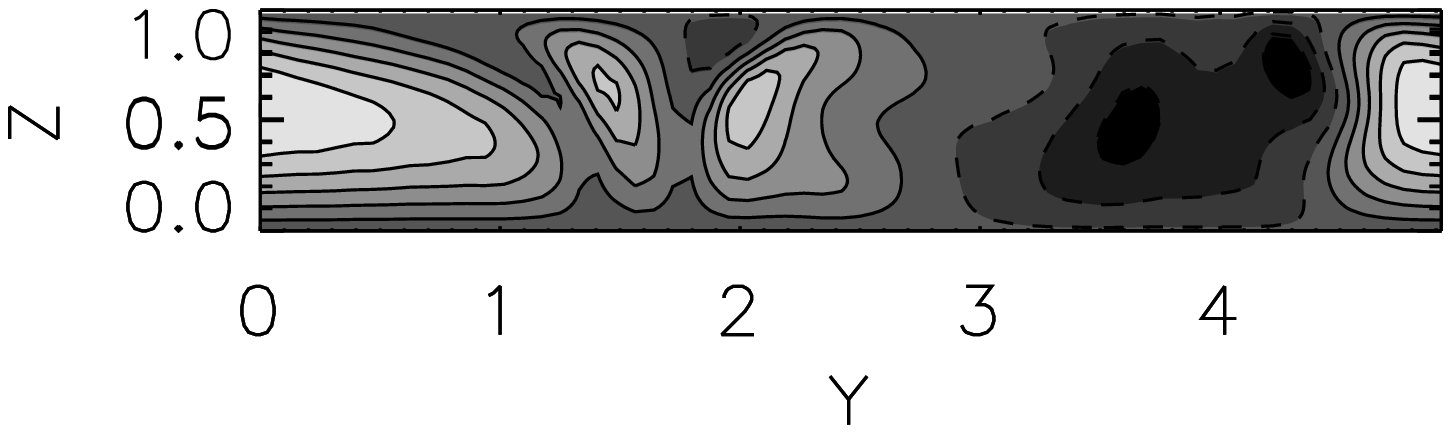,width=42cm}
\end{minipage}\hfill
\vskip -35.0cm
\begin{minipage}[t]{.35\linewidth}
\hskip -2cm \epsfig{figure=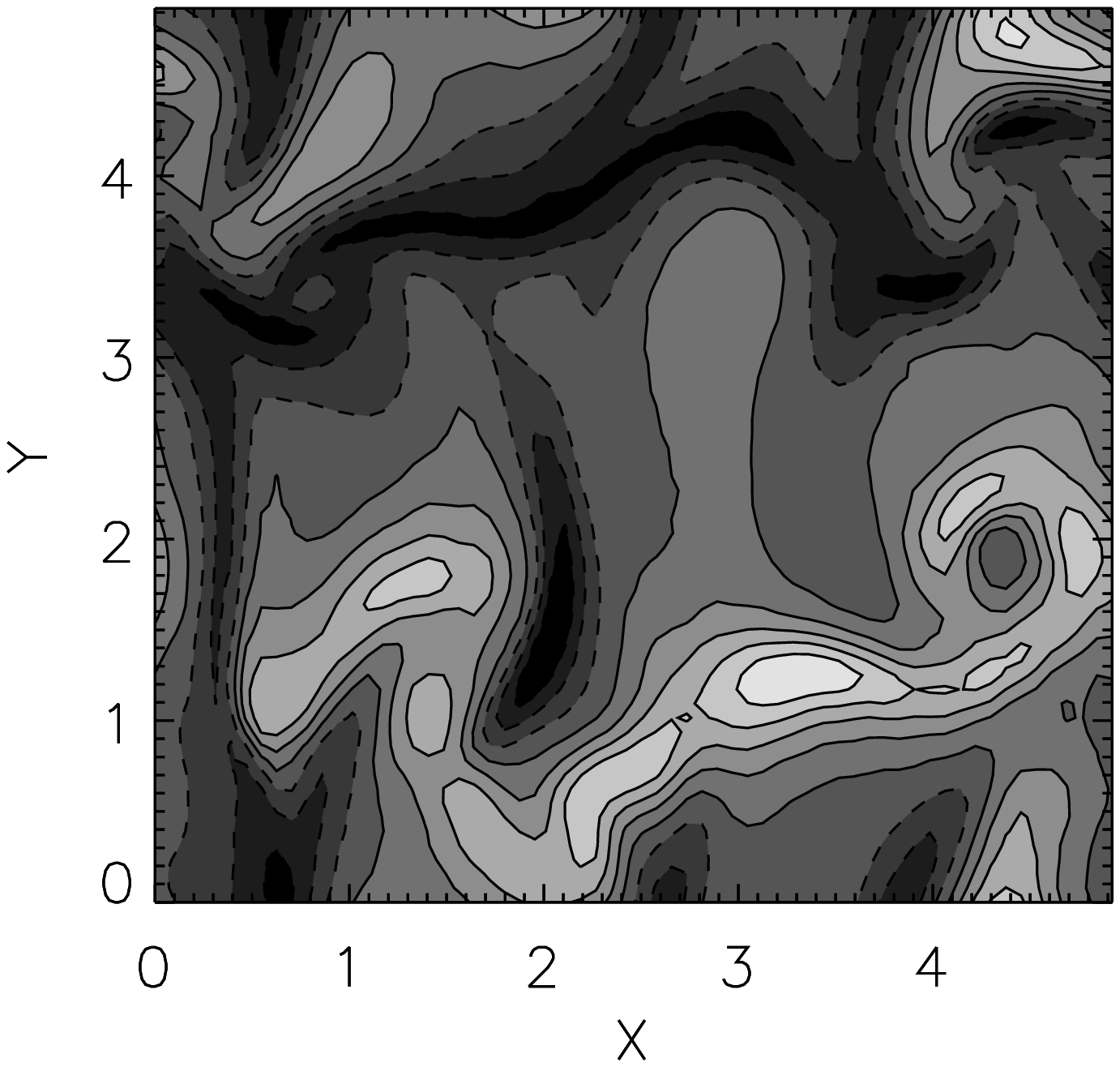,width=42cm}
\end{minipage}\hfill
\vskip -35.0cm
\begin{minipage}[t]{.35\linewidth}
\hskip -2cm \epsfig{figure=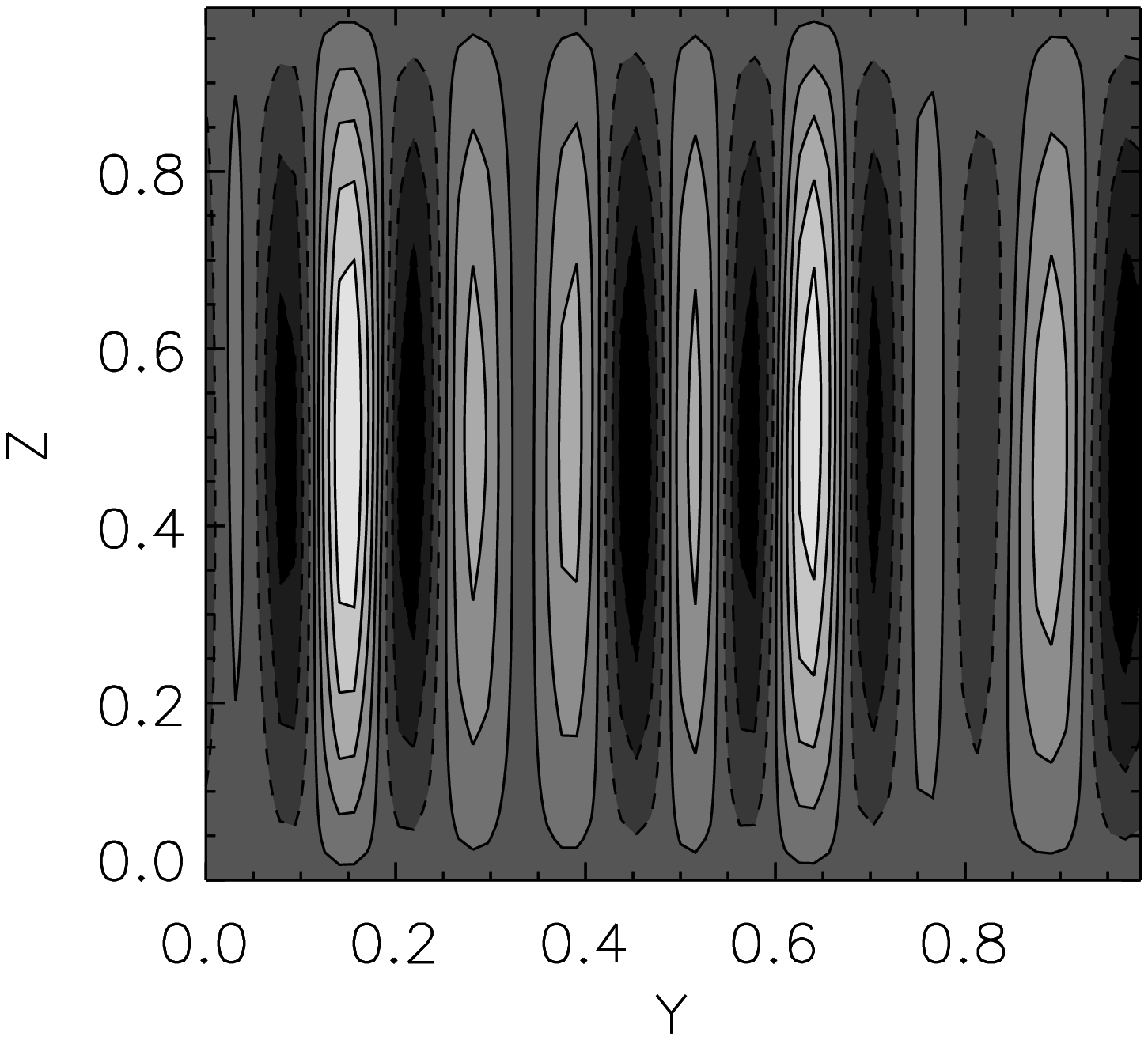,width=42cm}
\end{minipage}\hfill
\vskip -35.0cm
\begin{minipage}[t]{.35\linewidth}
\hskip -2cm \epsfig{figure=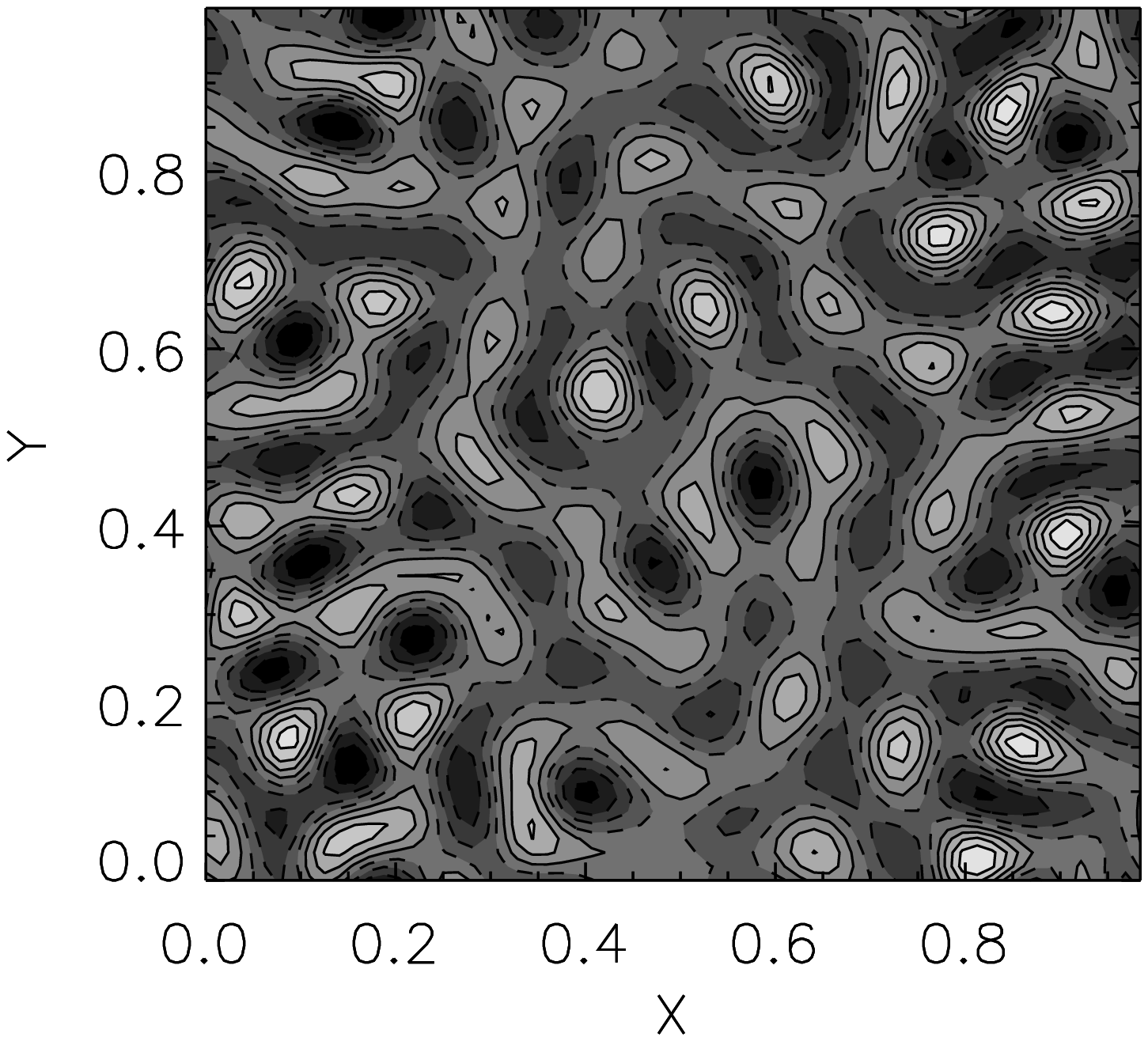,width=42cm}
\end{minipage}\hfill
\vskip 1.0cm
 \caption{
Sections of the vertical component of velocity,  $V_z$, without
rotation (regime  NR, two upper plots) at  $x=4.3$,
 and  $z=0.8$. Ranges of the filed:    $(-257,\, 506)$ and       $(-254,\, 572)$. Two lower plots demonstrate
 sections of    $V_z$ with rotation  (regime R1)  at  $x=4.3$ and
  $z=0.8$. Ranges of the filed:  $(-88,\, 127)$ and        $(-55,\, 86)$.
} \label{fig1}
\end{figure}

The onset of convection in the plane infinite layer is a
threshold  phenomenon which occurs with the increase of the
Rayleigh number to its critical value,
  $\Ra^{\rm cr}$.  Here we consider three regimes of convection.
 We consider simulations without rotation similar to \citep{Meneguzzi} (but without the magnetic field) and with rotation for two regimes
   with different
amplitudes of the heat sources. All simulations were carried out
using a quite rough grid, $\rm N=64^3$, which helped us to
present the statistics for the $T_3$ fluxes. Here are the regimes:
\begin{enumerate}
\item[NR:] Regime without rotation (the Coriolis term is dropped)~\footnote{As there is no rotation, $\Ra$ and $\E$ do not retain their physical meaning
defined in Section 2.1. More details on parametrization of
non-rotational magnetoconvection can be found in
\citep{Meneguzzi}.} , $\Ra=9\cdot 10^5$, $\Pr=1$, $\E=1$,
$\Re\sim 700$.

\item[R1:] Regime with rotation, $\Ra=4\cdot 10^2$, $\Pr=1$, $\E=2\cdot 10^{-5}$, $\Re\sim 200$.

\item[R2:] Regime with rotation, $\Ra=1\cdot 10^3$, $\Pr=1$, $\E=2\cdot 10^{-5}$, $\Re\sim 10^4$.

\end{enumerate}
\noindent Regime NR in Fig.\ref{fig1}
 corresponds to turbulent convection without rotation~\footnote{As at the onset of convection the horizontal scales
   $\rm L_x=L_y$
 of the convective cell are larger than the vertical scale  $\rm L_z$,
 the box for simulations is usually longer in $x$-,$y$-directions.  Here  $\rm L_x=L_y=5L_z\equiv 5L$.}
 with a quasi-periodic in time behavior of the kinetic energy   $E_K(t)$.
  Note that a small-scale hydrodynamic helicity appears $\dsize {\cal H}^{\cal H}={\bf V}\cdot {\rm rot }{\bf V}$.
 For the case without rotation, because of the absence of a preferred direction,  the mean helicity  $\dsize \overline{{\cal H}^{\cal H}_{\cal V}}$ is zero, see   Fig.\ref{zhel2}.  One can find the details of the problem in
  \citep{Moffatt,Kr,Zeldovich}. See also the results of DNS in \citep{Meneguzzi}.

Convection with rotation is characterized by the appearance  of
numerous vertical rotating columns (cyclones and anticyclones).
Their number is defined by the Ekman number as  $k_c\sim
\E^{-1/3}$ \citep{Ch,Roberts65,Busse70}, see also the review by
\cite{Jones} for details. For $\E\ll 1$ the number of cyclones
and anticyclones is the same, whereas in the solar-like regimes,
where the Coriolis and inertial terms are comparable, the
cyclones dominate \citep{Davidson}.  For the liquid core of the
Earth,  $\E\sim 10^{-15}$.  It  leads to $k_c\sim10^5$ which is
still impossible for  DNS.  Usually one is able to arrive at
  regimes with  $\E=10^{-4}\div 10^{-6}$   \citep{Jones}.  The purpose of  DNS is to find an asymptotic regime and then to  extrapolate the results of the simulations to the terrestrial parameters.
  The side effect of such cyclonic convection is the decrease of the energy scale of the system and, as a result,
  the  increase of viscous dissipation: the critical Rayleigh number  depends on the Ekman number as
  $\Ra^{\rm cr}\sim \E^{-1/3}$.

\begin{figure}[th!]
\vskip -33.0cm
\begin{minipage}[t]{.35\linewidth}
\hskip -2cm \epsfig{figure=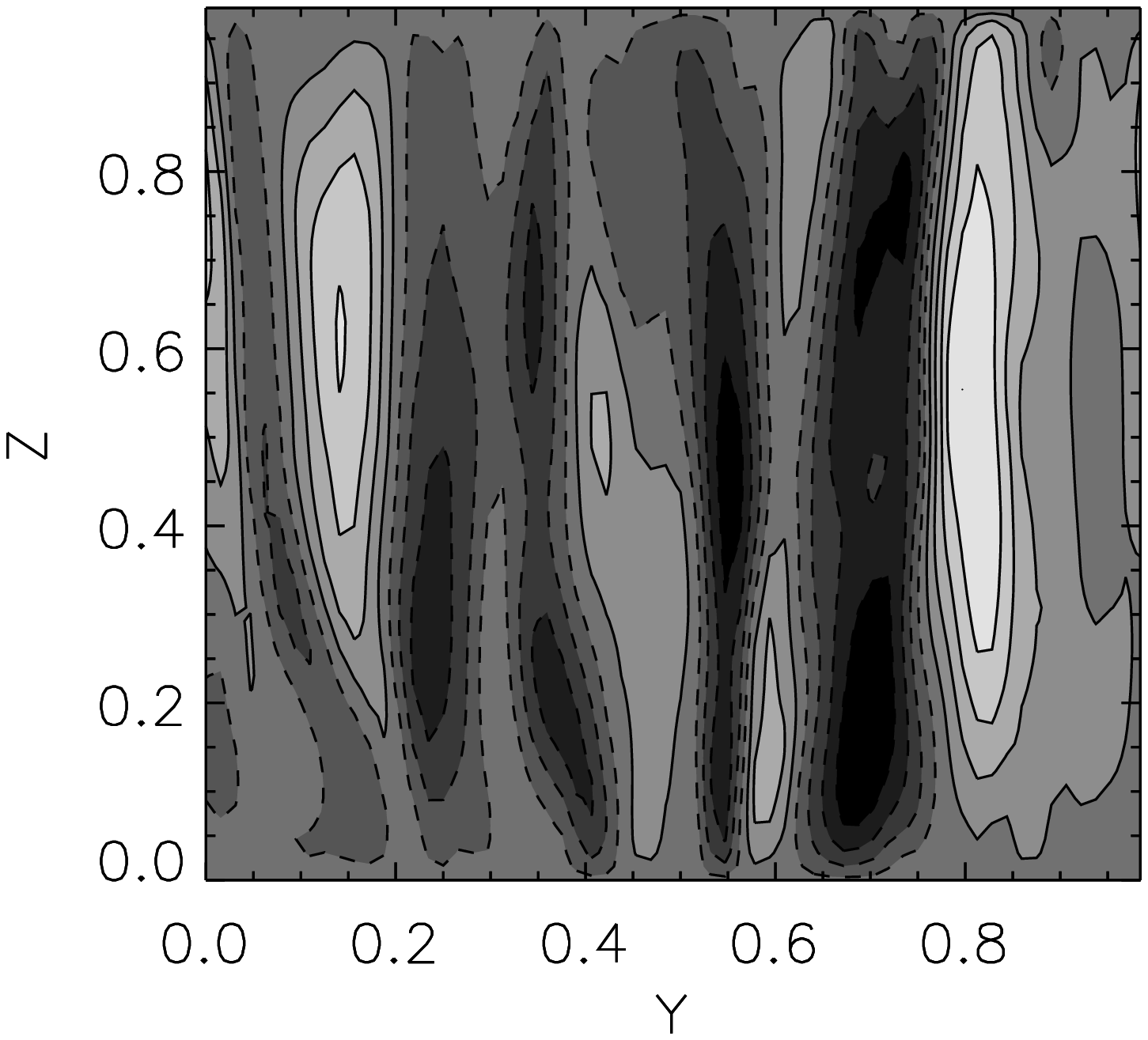,width=42cm}
\end{minipage}\hfill
\vskip -35.0cm
\begin{minipage}[t]{.35\linewidth}
\hskip -2cm \epsfig{figure=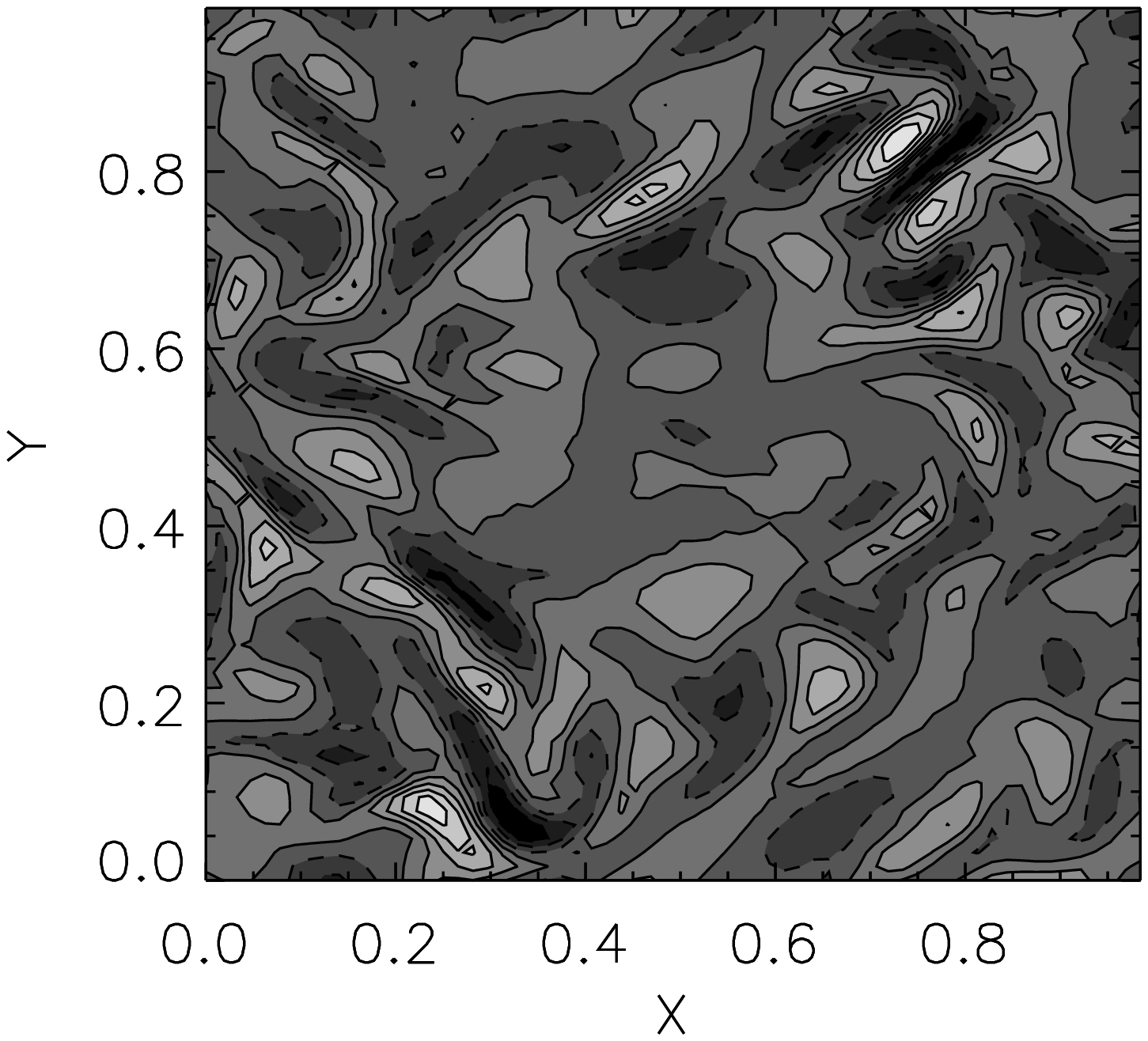,width=42cm}
\end{minipage}\hfill
\vskip 1.0cm
 \caption{
Sections of the vertical component of velocity  $V_z$ with
rotation (regime  R2) at  $x=4.3$,
 and  $z=0.8$. Ranges of the filed:    $(-407,\, 449)$ and       $(-456,\, 600)$.
} \label{fig1a}
\end{figure}

\begin{figure}[th!]
\psfrag{EAS}{ $\dsize {\cal H}^{\cal H}$} \hskip 0cm
\epsfig{figure=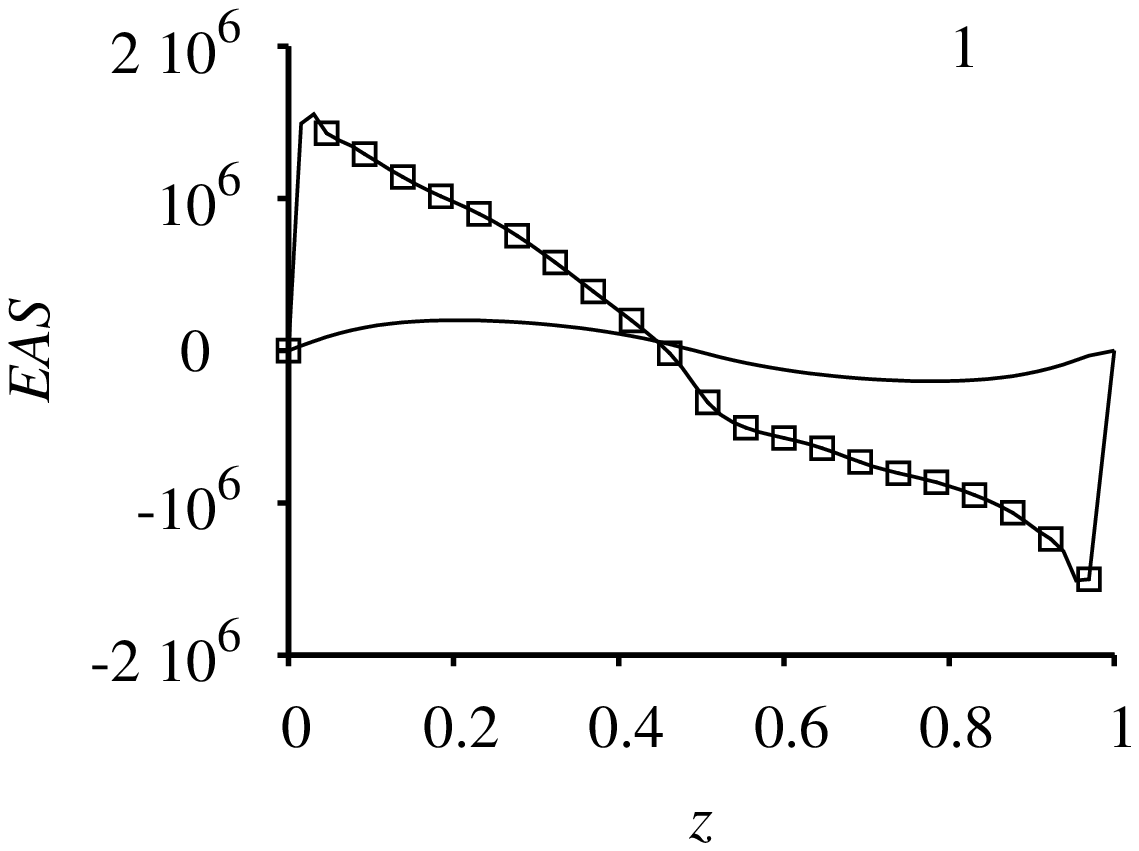,width=8cm}
%
\hskip 0cm \epsfig{figure=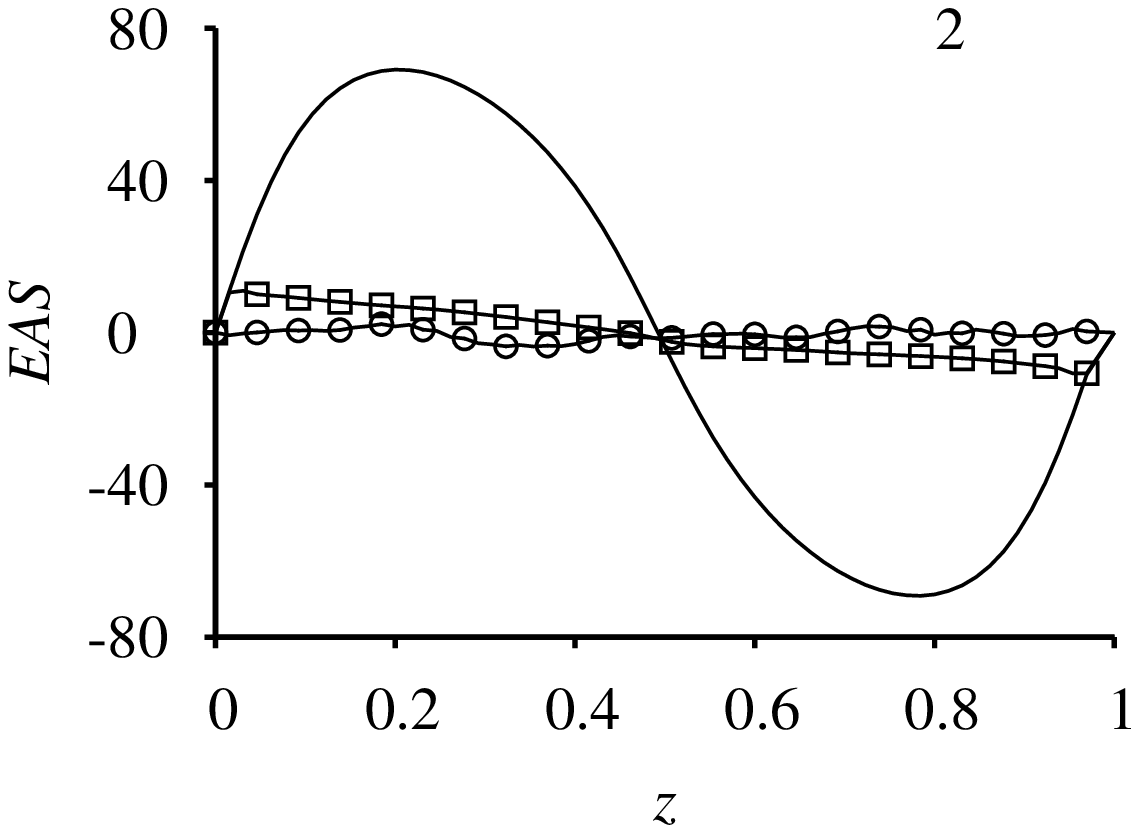,width=,width=8cm}
\vskip 1cm
 \caption{
The profiles of the hydrodynamic helicity  $\dsize {\cal H}^{\cal
H}$
   for
$\E=3 \, \cdot 10^{-5}$, $\Pr=1$, $\Ra=4\cdot 10^2$ (solid line)
  and  $\E=3 \, \cdot 10^{-5}$, $\Pr=1$,  $\Ra=1.2\cdot 10^3$ (squares)  (1).  Values are normalized
 at the mean over the volume kinetic energy
  ($\overline{E_K}$)  (2). Circles correspond to the regime NR.
} \label{zhel2}
\end{figure}

Regime R1 corresponds to the geostrophic state near the onset of
convection, see Fig.\ref{fig1}.
 The increase of  $\Ra$ (regime R2) leads to the suppression of the regular cyclones and appearance of the small scale flows in the $z$-direction,
 and deviation from the geostrophic state to the so-called quasi-geostrophic turbulence, see
  Fig.\ref{fig1a}.  The nonlinear term tends to the amplitude of the Coriolis force and pressure gradient,
  the temporal behavior of the system becomes chaotic.  In both the cases (R1, R2) the mean non-zero helicity generates:
  $\dsize {\cal H}^{\cal H}=<{\bf V}\cdot {\rm rot} {\bf V}>_{xy}$ (the average is taken over the $(x,\,y)$ plane),
 see Fig.\ref{zhel2}.   As was mentioned before,    $\dsize {\cal H}^{\cal H}$ is zero for regime NR.
For R2, $\dsize {\cal H}^{\cal H}$ is close to the linear
function in the main volume.

\begin{figure}[th!]
\vskip -3.0cm
\psfrag{PSI}{$E_K$} \hskip 2cm
\epsfig{figure=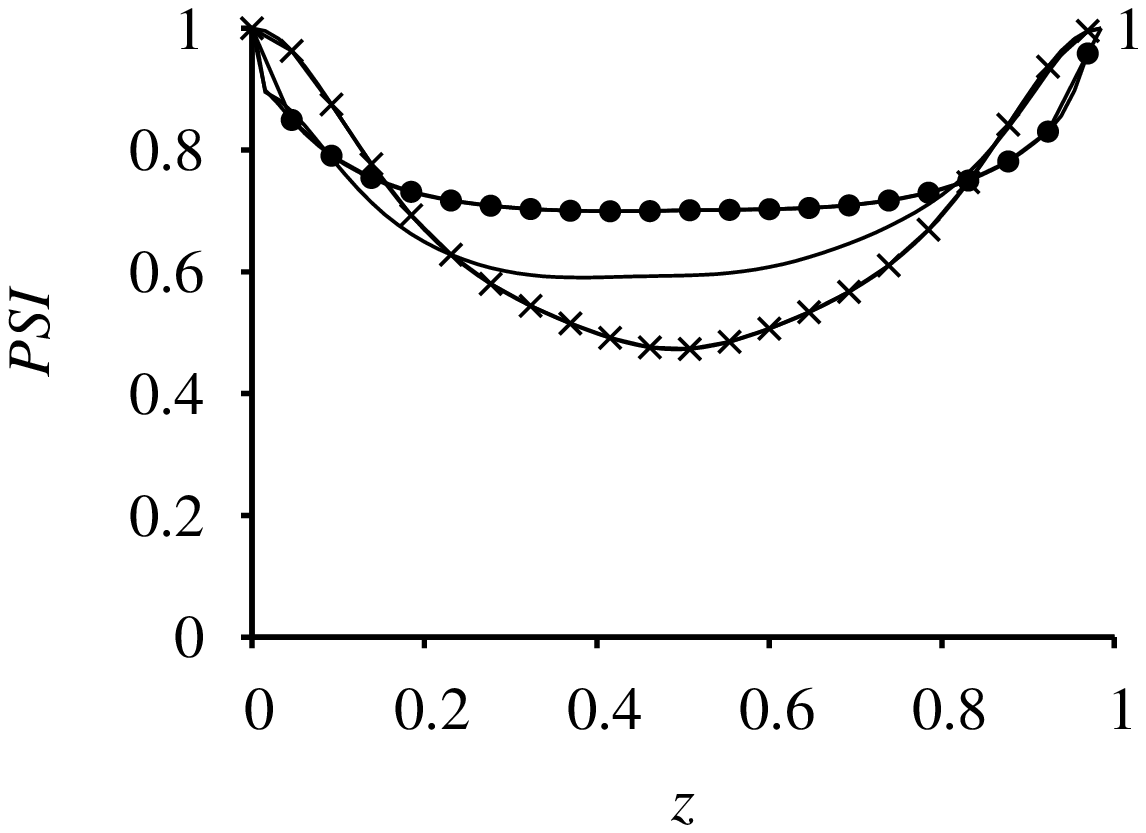,width=,width=10cm}
\vskip -1.0cm
\psfrag{PSI}{$E_K^\perp$} \hskip 2cm
\epsfig{figure=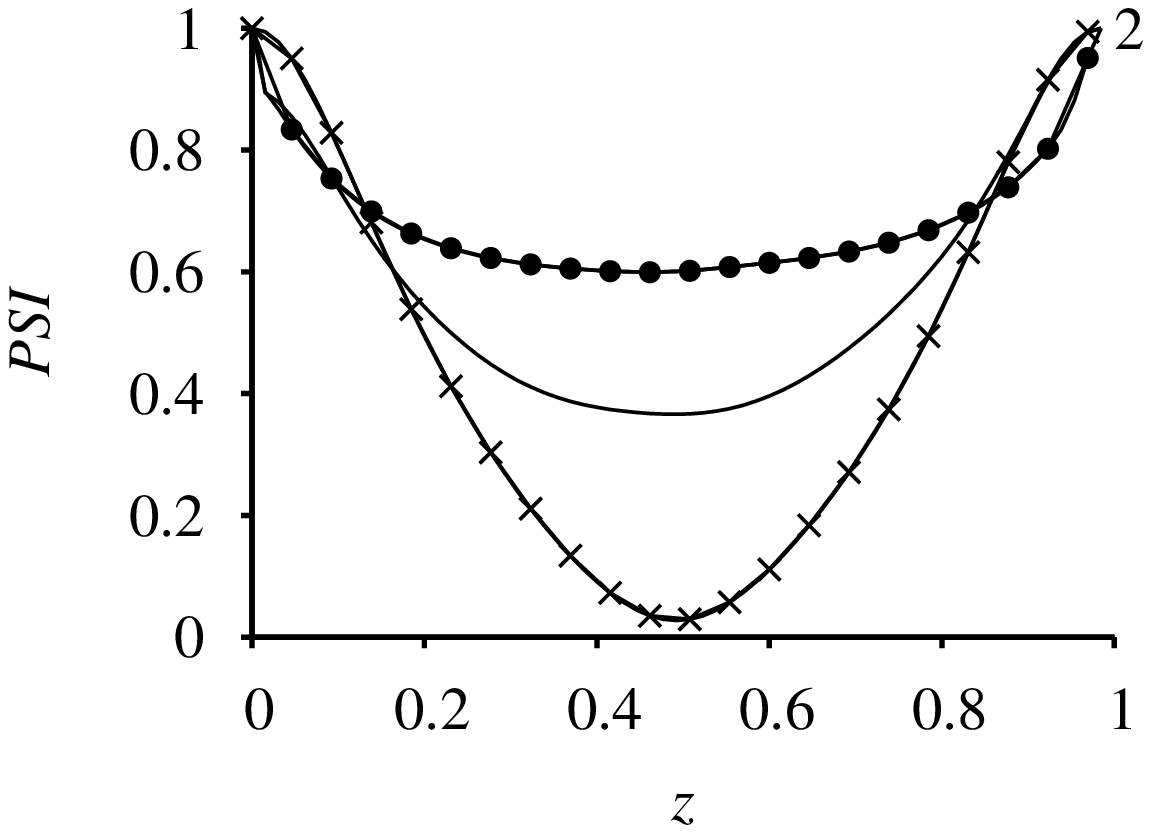,width=,width=10cm} \vskip -1.0cm
\psfrag{PSI}{$E_K^{||}$} \hskip 2cm
\epsfig{figure=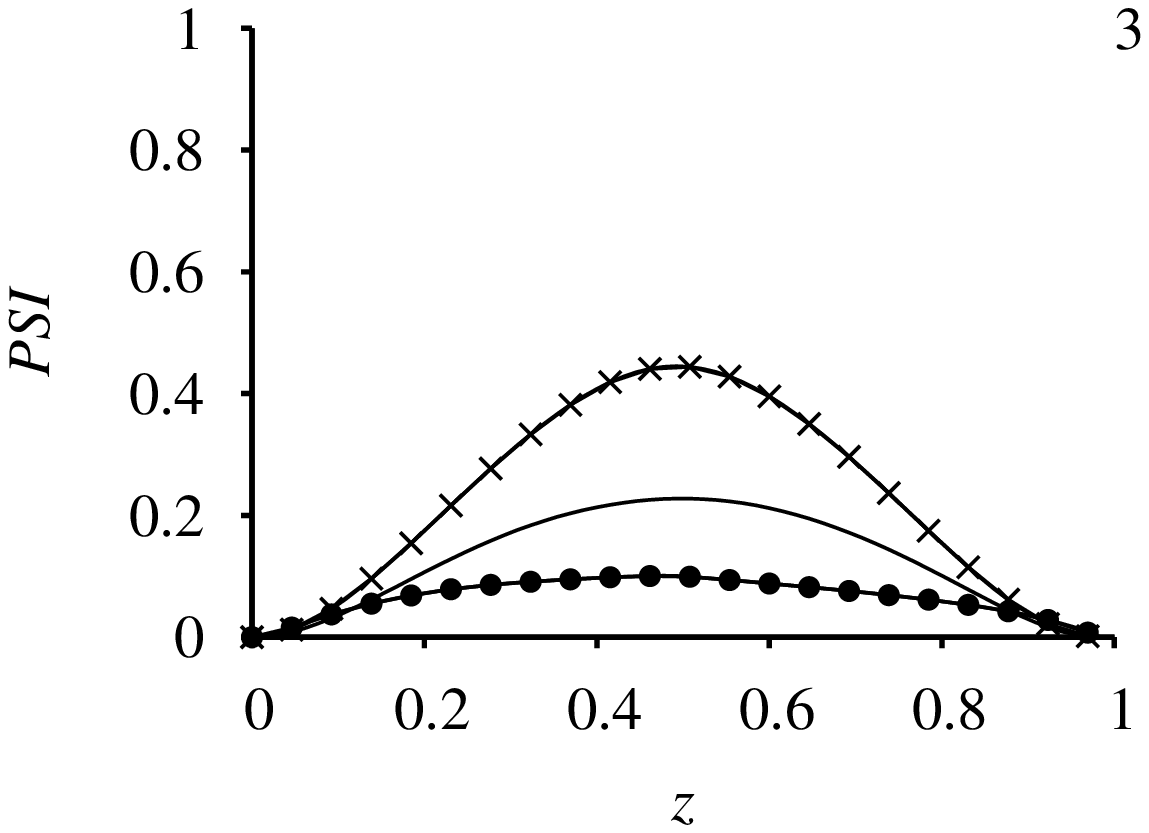,width=,width=10cm} \vskip -1.0cm \vskip
1.0cm
 \caption{
Dependence of the  full kinetic energy $E_K$ (1), transverse  $
E_K^{\perp}$ (2) and
 longitudinal
$ E_K^{||}$ (3) components for regimes  NR (solid line), R1
(crosses) and R2 (diamonds). All quantities are normalized at
$E_K$. } \label{fig3}
\end{figure}

 It is instructive to analyze the behavior of the kinetic energy as a function of depth for different components of the velocity field,
  see Fig.\ref{fig3}.   The different behavior of the transverse
$ E_K^{\perp}=\left(V_x^2+V_y^2\right)/2$ and
 longitudinal $ E_K^{||}=V_z^2/2$ components of the energy  depends on the different boundary conditions. For the nonrotating case NR,
  the values of the components are comparable:
$\dsize {\lambda }_{\rm NR}=
 {E_K^{\perp}\over 2 E_K^{||} } {\Big{|}_{z=0.5}}
\sim 1$,  where the factor  2  appears because of the summation
over the two horizontal directions. In contrast, for regime R1
 in the vicinity of the central part of the volume, the
 transverse velocity changes its sign and, as a result, the ratio increases: $\dsize {\lambda }_{\rm R1}\sim 0.1$.
 This reflects the symmetry property of the flow at the threshold of generation. It is interesting that the increase of  $\Ra$ (regime R2) leads to
reinforcing the anisotropy in the whole volume in the opposite
direction: $\dsize {\lambda }_{\rm R1}\sim 5$, i.e. at the rather
large  $\Ra$  and under strong rotation, vertical motions are
suppressed with rotation. This is the so-called degeneration of
three dimensional convection to the two-dimensional predicted by
\citep{Batch}, see also the similar degeneration caused by a
strong mean magnetic field \citep{Kraichnan65}.  Also, the
increase of  $\Re$ (R2)
  goes with the appearance of the layer structures at  $z=0,\, 1$. The latter  is connected with the formation of a thermal boundary layer, thickness  $\delta_T\sim \Ra^{-1/3}$.

\vskip 0.5cm
\section{Spectra of the fields and fluxes in the wave space} \label{p4}
\begin{figure}[th!]
\vskip -4.0cm \psfrag{PSI}{$k$} \psfrag{S}{$E_K$} \hskip 2cm
\epsfig{figure=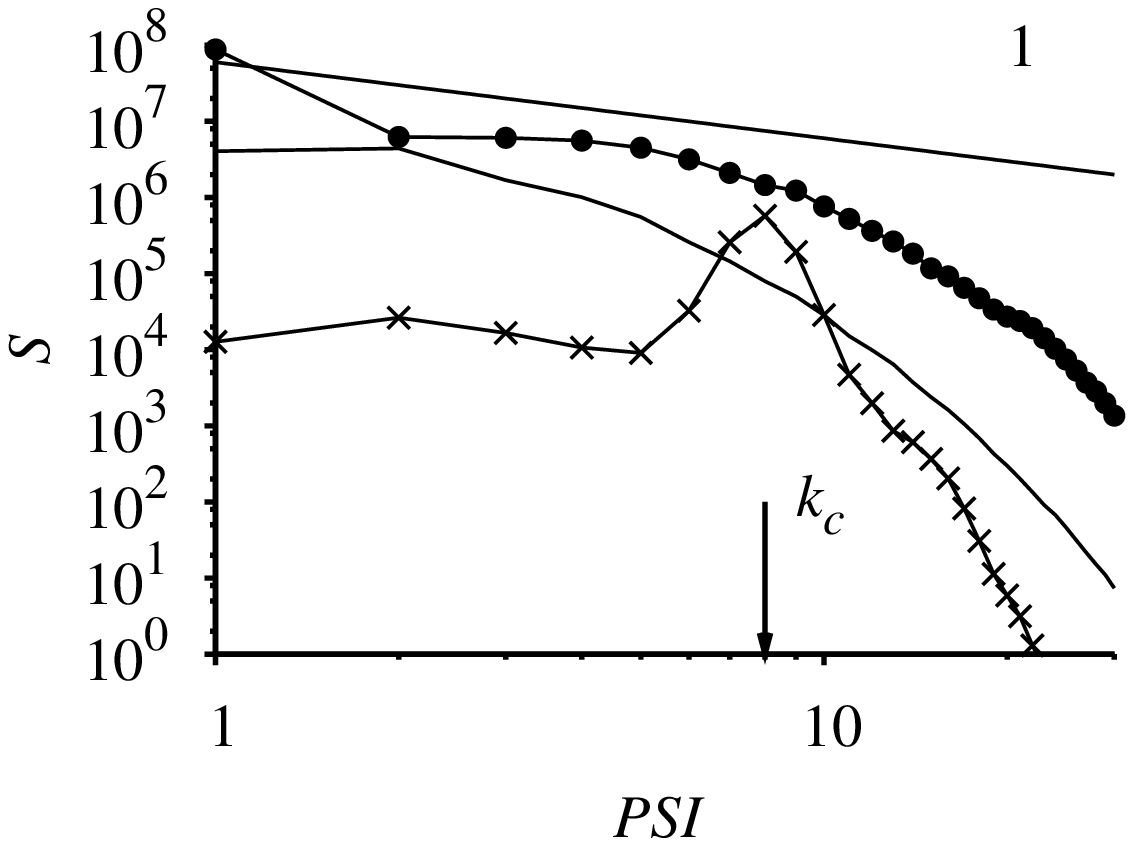,width=,width=10cm}\vskip -1.0cm
\psfrag{PSI}{$k_\perp$}\psfrag{S}{$ E^\perp_K$} \hskip 2cm
\epsfig{figure=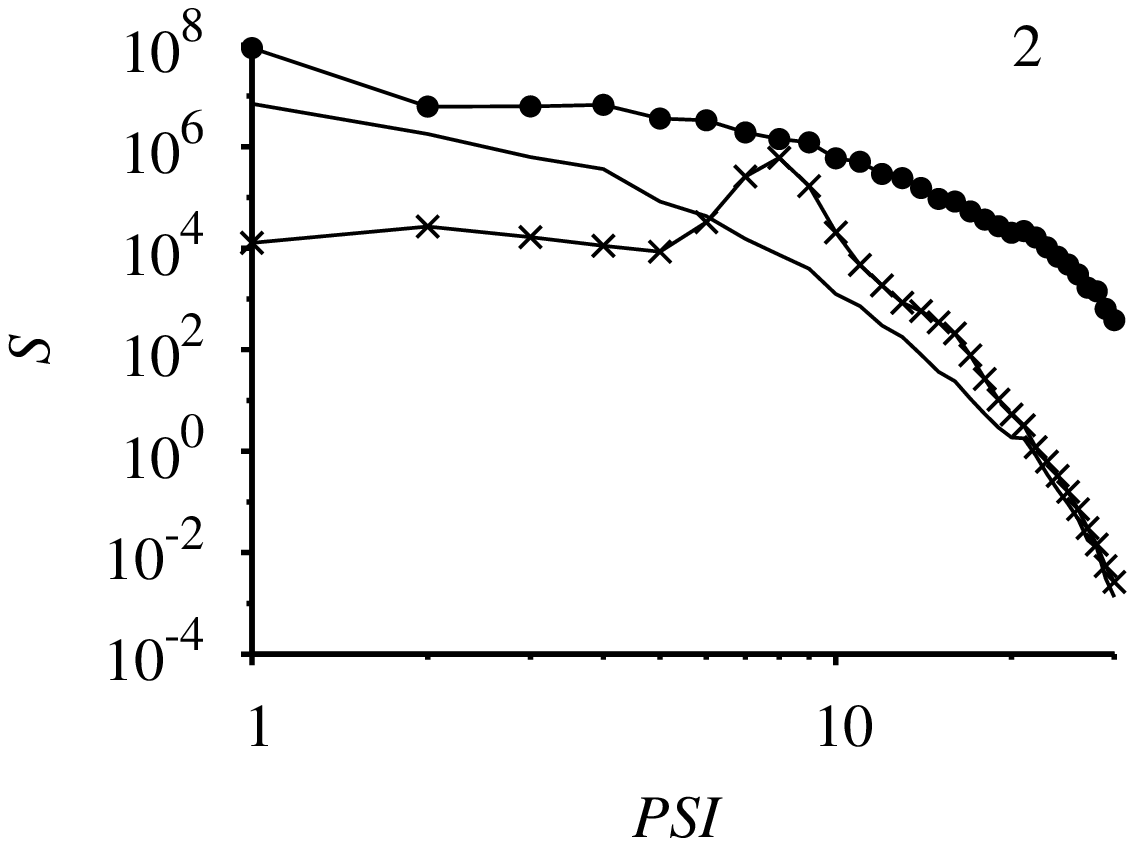,width=,width=10cm}\vskip -1.0cm
\psfrag{PSI}{$k_{||}$}\psfrag{S}{$E^{||}_K$} \hskip 2cm
\epsfig{figure=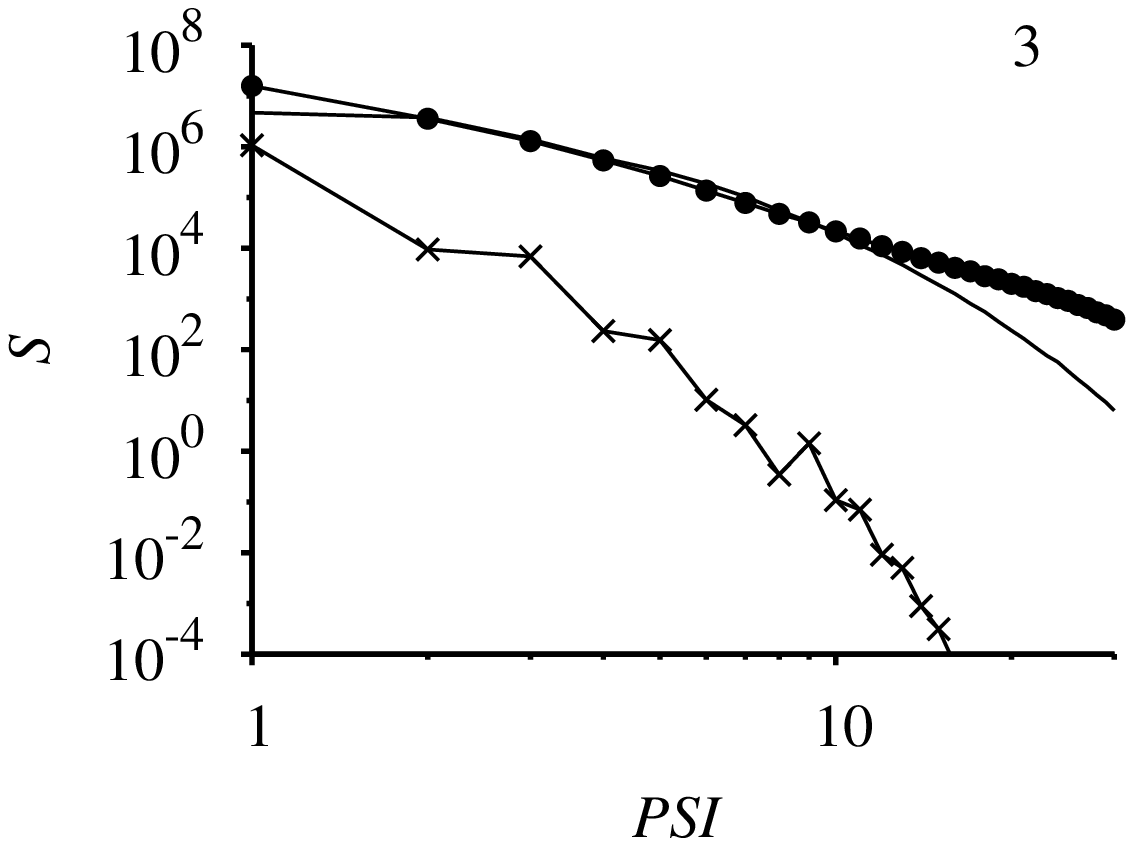,width=,width=10cm}
\vskip -0cm
 \caption{Spectra of kinetic energy  $ E_K$,   $ E_K^{\perp}$,  $E_K^{||}$  as a function of $k$ (1), $k_\perp$ (2) and
 $k_{||}$ (3) components for regimes  NR (solid line), R1 (crosses) and R2 (circles).  The straight line
 corresponds to the Kolmogorov's law of   $\sim k^{-5/3}$.
} \label{fig4}
\end{figure}

 Here we continue our analysis of the fluxes in k-space performed in \cite{HR08} and consider anisotropy effects.
For our regimes, we shell consider the integral spectrum  of
kinetic energy  $\dsize E_K({k})$,  as well as its
 longitudinal
$\dsize E_K^{||}({ k_{||}})=\int\int E_K({ k_x, k_y, k_z})\, { d
k_x d k_y}$
 and transverse
$\dsize E_K^\perp({ k_\perp})=\int E_K({ k_x, k_y, k_z})\, { d
k_z}$ spectra, where $ k_\perp^2=k_x^2+k_y^2$, ${ k_{||}}\equiv
{  k_z }$. For regime  NR, the spectral estimates are close to
the Kolmogorov dependence  $ \sim k^{-5/3}$, see Fig.\ref{fig4}.
  In other words, the spectra for NR are isotropic. It corresponds to the isotropic form of the convective cells at the main scale of the box.

The spectra of convection with rotation differ from those for NR.
 Regime R1 is close to the onset of convection. The integral spectrum demonstrates a well-pronounced maximum,
 which corresponds to the scale of cyclones $ k_c\sim 8$.  The increase of $\Ra$ (R2) fills the gap for $k<k_c$ and the spectrum starts to resemble the spectrum without rotation.

 Spectra   $\dsize E_K^\perp$ and
$\dsize E_K^{||}$ for R1 are different:   spectrum  $E_K^{||}$
does not feel boundary $ k=k_c$, whereas $\dsize  E_K^\perp$ at
$ k<k_c$ is close to the white noise, and at large  $ k_\perp$
the spectrum decays.
 For larger $\Ra$,  $\dsize E_K^{||}$ tends to Kolmogorov's asymptotic, and the transverse spectrum  is still white for small $k$ and
 close to the one without rotation at  $ k_\perp>k_c$.

Note that for R1 and R2 the form of the integral spectrum $\dsize
E_K$ is  defined by the form of  its transverse component
 $\dsize E_K^\perp$.

The apparent similarity in the spectra for NR and R2 does not
mean that the physics of the processes are similar. In this
respect we remind the reader that the  two-dimensional turbulence
with inverse cascades and the three-dimensional turbulence with
direct cascades have the same slope of "-5/3" for the kinetic
energy spectrum \citep{KM}. However, the  directions of the
energy transfer through the spectrum in these two examples are
opposite.

\begin{figure}[th!]
\vskip -4.0cm \psfrag{k1}{$k$} \psfrag{T1}{${T_K}$} \hskip 2cm
\epsfig{figure=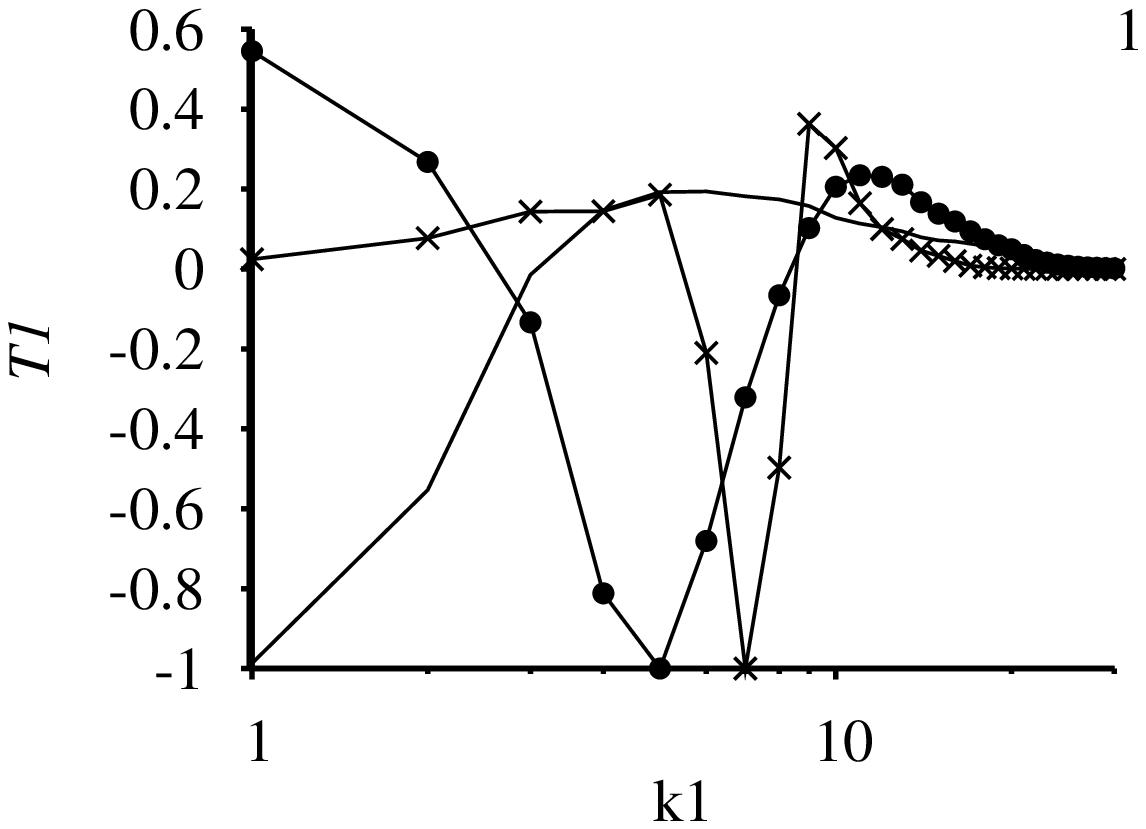,width=,width=10cm}\vskip -1.0cm
\psfrag{k1}{$k$} \psfrag{T1}{$T_K^\perp$} \hskip 2cm
\epsfig{figure=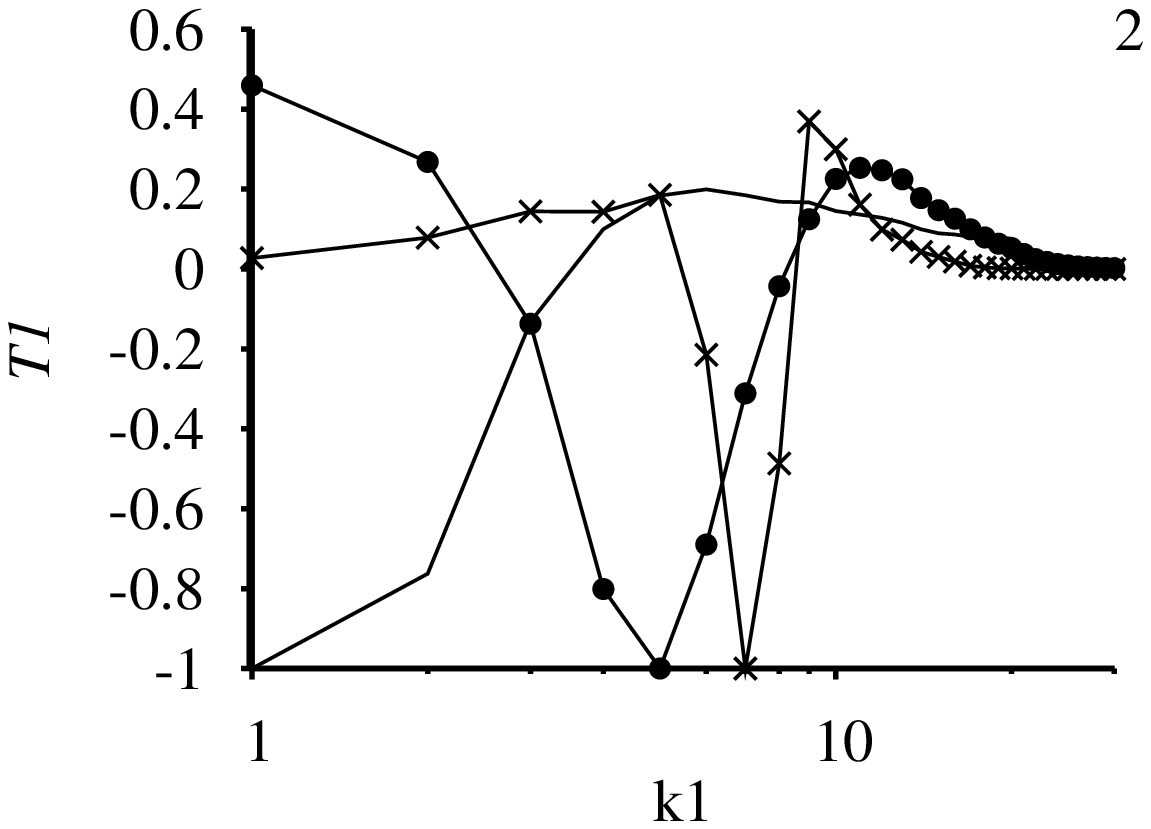,width=,width=10cm}\vskip -1.0cm
\psfrag{k1}{$k$} \psfrag{T1}{$T_K^{||}$} \hskip 2cm
\epsfig{figure=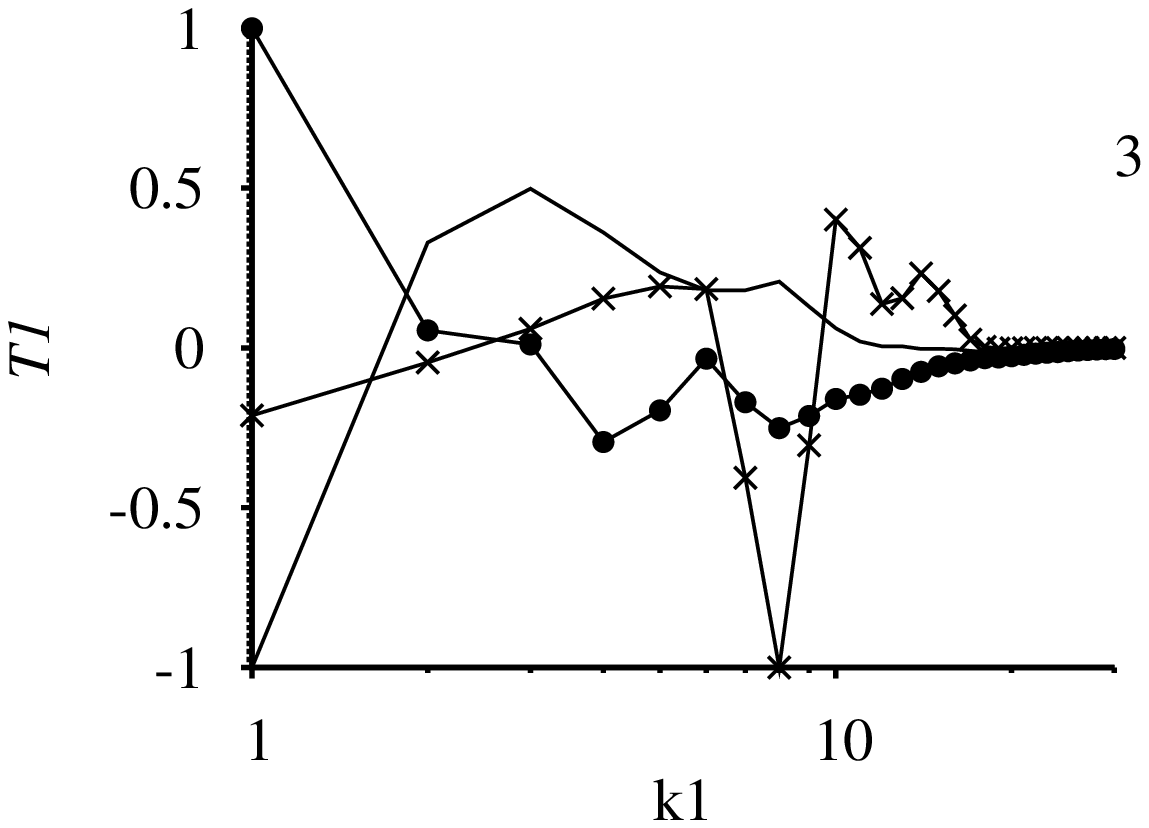,width=,width=10cm}
 \caption{ The fluxes of kinetic energy  $T_K$ (1),  $T_K^\perp$ (2), $T_K^{||}$ (3)
  as a function of  $ k$  for regimes
  NR (solid line), R1 (crosses) and R2 (circles). Functions are normalized at extremums of  $T_K$.
} \label{fig5}
\end{figure}

Fig.\ref{fig5} shows the fluxes of the kinetic   ${ T}_K$
energies for the regimes mentioned above. At first we consider
the integral over all directions.
 Regime  NR  for  ${ T}_K$ demonstrates the well-known behavior for the direct Kolmogorov cascade in 3D.
 For large scales
   ${ T}_K<0$, these scales are donors and provide energy to the system. On the other hand, the harmonics with
   large $k$ absorb energy. The two-dimensional turbulence exhibits  mirror-symmetrical behavior relative to
    the axis of the abscissa  \citep{KM}. In this case the energy cascade is inverse.

Rotation  essentially changes the behavior of the fluxes of
kinetic energy.  The leading order wave number is
   $k_c$. For  $k>k_c$ we also observe the direct cascade  of energy ${ T}_K>0$. The maximum of
  ${ T}_K$ is shifted relative to the maximum of the energy to large $k$;
  the larger $\Re$, the stronger the shift.
    For  $k<k_c$, the behavior is more complex: for
 small $k$, the inverse cascade of kinetic energy takes place,  ${ T}_K>0$. It is very tempting to associate this small region
 with the appearance  of coherent structures, see \citep{Tabel}.
 On the other hand, for the larger region of $k$
 $(0\dots k_c)$ we still have the direct cascade ${ T}_K <0$. The increase of
  $\Re$  leads to the narrowing of the region with the inverse cascade and to the increase of the inverse flux. One may
   suggest that the change of the sign of flux
 ${ T}_K$ at  $k<k_c$ is connected with the appearance of the non-local energy transfer:
   so that the energy to the large-scales  ${\bf k_1}$
  comes from modes
$|{\bf k_2}|\sim |{\bf k_3}|  \gg |{\bf k_1}|$,
  ${\bf k_1}={\bf k_2}+{\bf k_3}$ \citep{Waleffe}.  Hence, in the case of rotation, two cascades of kinetic energy (direct and inverse)
     take place simultaneously.  As shown in the simulations for the
  higher resolution \citep{HR08}  these results do not depend on the presence of the magnetic field, if the kinetic and magnetic energies are comparable in order of magnitude.

 To estimate the anisotropy of the fluxes, we decompose  ${ T}_K$  into the sum of  the transverse ${ T}_K^{ \perp}$ and
 longitudinal  ${ T}_K^{ ||}$ parts, so that in  Eq.(\ref{sys100a}) the summation over index  $j$
 was used for $j=1\dots 2$ and $j=3$, respectively.
  For all three regimes, the fluxes in the z-direction are much smaller than in the horizontal plane, however, the form of ${ T}_K^{ ||}$ are similar to that of ${ T}_K$.
  This asymmetry can be explained as follows. Obviously, flux
 $\dsize \Pi_K^{||}\sim V_z^< {\partial E_K\over \partial z}$ is small for geostrophic
 flows.
 For NR, it is also small because the kinetic energy does not change too much along the length of the plumes in the z-direction in the main volume.
 As regards the boundaries $z=0,\, 1$, $V_z$ is small there. That is why the main contribution to  $\Pi_K$ yields the transverse flux:
$\dsize \Pi_K^\perp\sim {\bf V_\perp^<}\cdot \nabla_\perp E_K$.
 This asymmetry distinguishes all three regimes from the homogeneous, isotropic models.

 \section{Locality of the energy transfer} \label{p5a}
\begin{figure}[th!]
\vskip -3.0cm \psfrag{K}{ $  {K}$} \psfrag{Q}{ $  {Q}$} \hskip
-5cm \epsfig{figure= 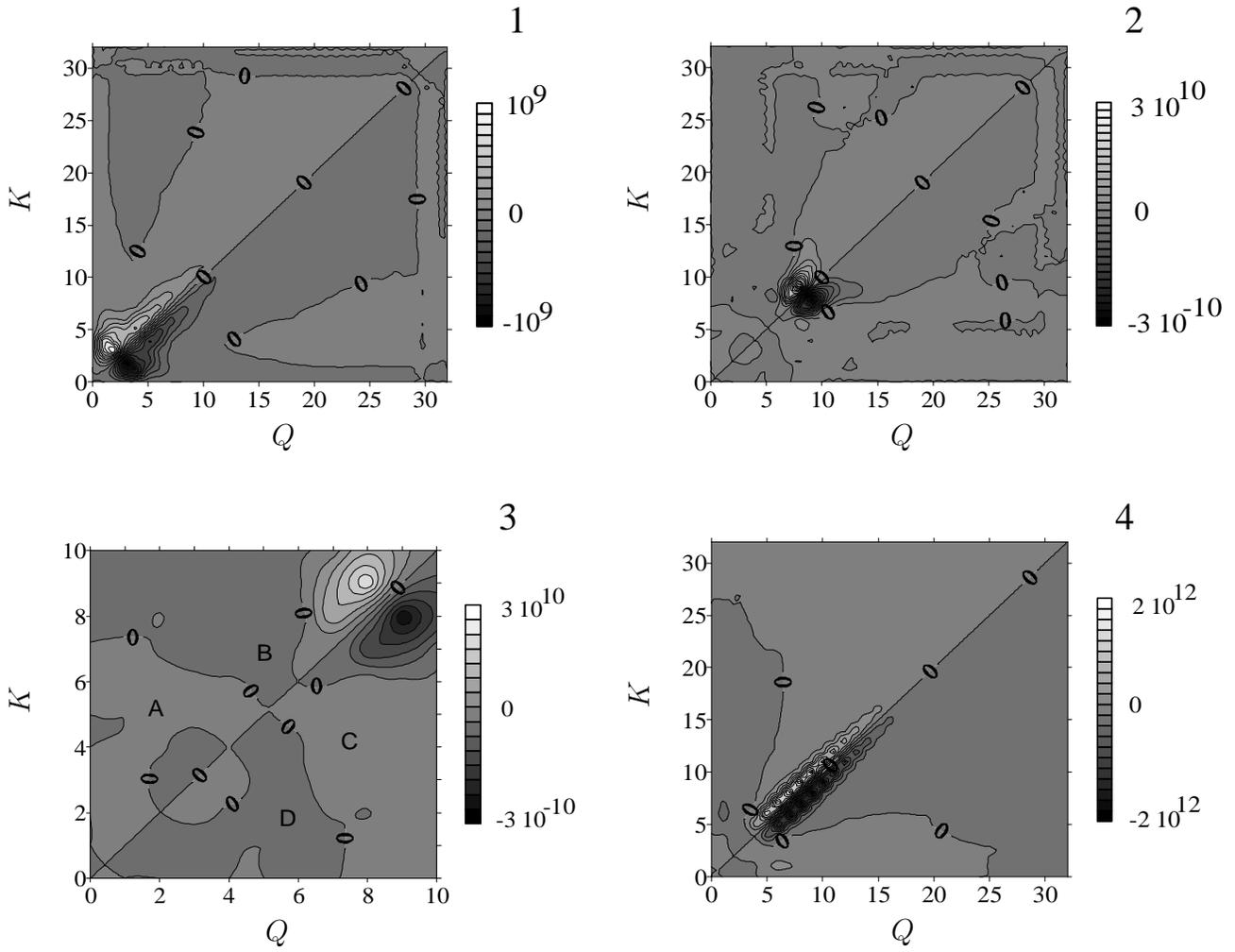,width=25cm} \vskip -10.0cm
 \caption{ The fluxes of kinetic energy ${ T}_{2}(k)$  for regimes NR (1), R1 (2 and 3), R2 (4).
} \label{fig6}
\end{figure}

\begin{figure}[th!]
\vskip -4.0cm
\begin{minipage}[t]{.35\linewidth}
\hskip -0.0cm \epsfig{figure=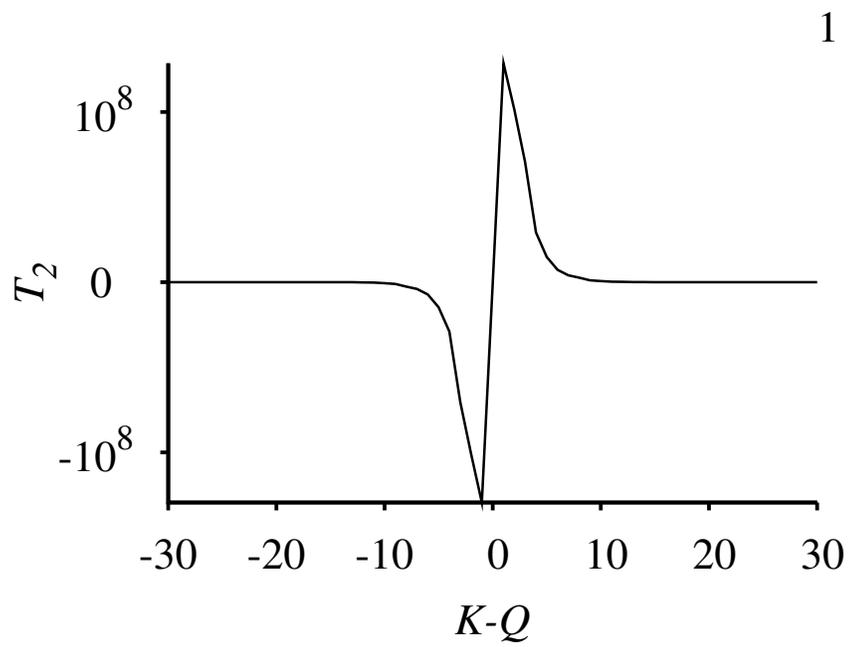,width=12cm}
\end{minipage}\hfill
\vskip 0.0cm
\begin{minipage}[t]{.35\linewidth}
\hskip  0.0cm \epsfig{figure=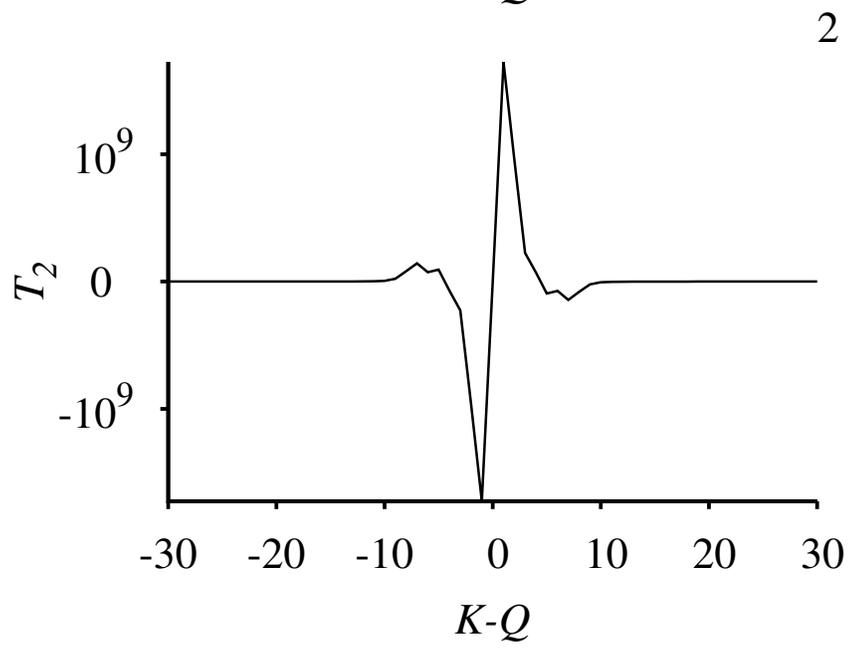,width=,width=12cm}
\end{minipage}\hfill
\vskip 0.0cm
\begin{minipage}[t]{.35\linewidth}
\hskip 0.0cm \epsfig{figure=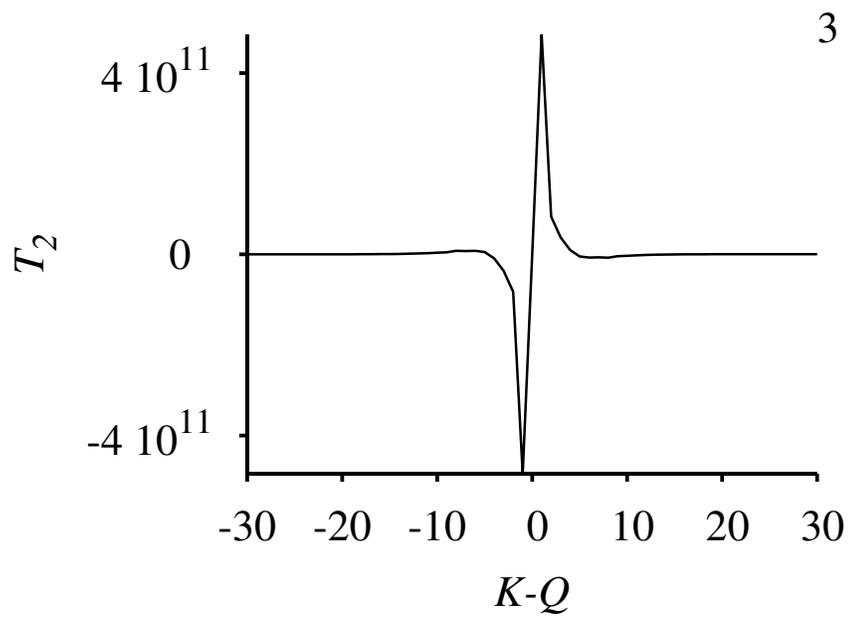,width=12cm}
\end{minipage}\hfill
\vskip 1.0cm
 \caption{   The fluxes of kinetic energy ${ T}_{2}(K-Q)$  for regimes NR (1), R1 (2), R2 (3).
} \label{fig7}
\end{figure}

 Let us consider the structure of the triad interactions. Fig.\ref{fig6} is a diagram of the antisymmetric (with respect to the diagonal
  $K=Q$) fluxes  ${ T}_{2}$ for regimes NR, R1, R2.    For NR, the results are similar to the case with the imposed force in \citet{ALEX07}:
 harmonics with  $K>Q$ take the energy from harmonics with  $ K<Q $ (direct cascade of the energy).
 The maximum of the energy flux corresponds to the  closest to the diagonal harmonics with   $K\sim Q$,
 i.e. the local energy transfer exists. Note that there are regions (i.e.  $Q=5$, $K=20$),
 in which there is a nonlocal inverse cascade of energy. Omitting these details, the whole behavior is close to the idealized Kolmogorov scenario.
 As a matter of convenience, we present the diagrams  as a function of
  $K-Q$ in Fig.\ref{fig7}.  This clearly demonstrates the existence of the direct cascade, as well as the local interactions with the local energy transfer.

The rotation changes the behavior of  $T_{2}$  at $k<k_c$,
 leaving it unchanged  at  $k>k_c$. We shall consider it in more detail. The positions of the extrema are near to
 $k_c$, where the energy transfers to the region of  $k$  larger than the leading mode $k_c$. On the other hand, at
  $k<k_c$ the flux decreases, which corresponds to the approach to  the state of statistical equilibrium, observed in Fig.\ref{fig5}.
  On the larger scale (see Fig.\ref{fig6}(3))  we resolve the finer structure at small  $k$:
 region A with a direct cascade of energy, but with an equiprobable energy transfer from the small  $Q\sim K$,
  as well as from the rather large  $Q\sim 4K$.   Region  B
 with an inverse cascade (as well as in region A with the weak energy flux of the order of 1/10 of its maximal absolute value)
 has an elongated strip-like  form  from  $Q\sim K$ to  $Q\sim 10 K$.  This corresponds to the occurrence of the small negative minimum at  $K>Q$~\footnote{We
 do not consider the behavior of function  ${ T}_{2}$ at  $K<Q$ (regions C and D), because of
 its antisymmetry.}.

The increase of   $\Ra$ (regime R2) leads to the shift of the
region with the inverse cascade in the direction of the small
$Q$ and  $K>Q$. As before, we observe the equilibrium state at
$k<k_c$. This regime exhibits a longer interval of the wave
numbers at
  $k>k_c$  with local energy transfer and the direct cascade. The relative input of the region with the inverse cascade becomes smaller, see Fig.\ref{fig7}(3)
 and  at the same time this input shifts to the large-scale field at  $k\ll  k_c$,
 which can be interesting for geodynamo applications, where  $k_c\sim 10^{5}$
 and the geomagnetic field generation regions ($k\sim 1\div 10^3$  for the typical magnetic Reynolds number  $\Rm\sim 10^2\div 10^3$)
 are separated at least  by some orders of magnitude of $k$.

\vskip 0.5cm
\section{Locality of interactions} \label{p5}
\begin{figure}[th!]
\vskip -4.0cm
\begin{minipage}[t]{.35\linewidth}
\psfrag{P}{ \tiny $P$} \psfrag{Q}{ \tiny $Q$} \hskip -0.0cm
\epsfig{figure=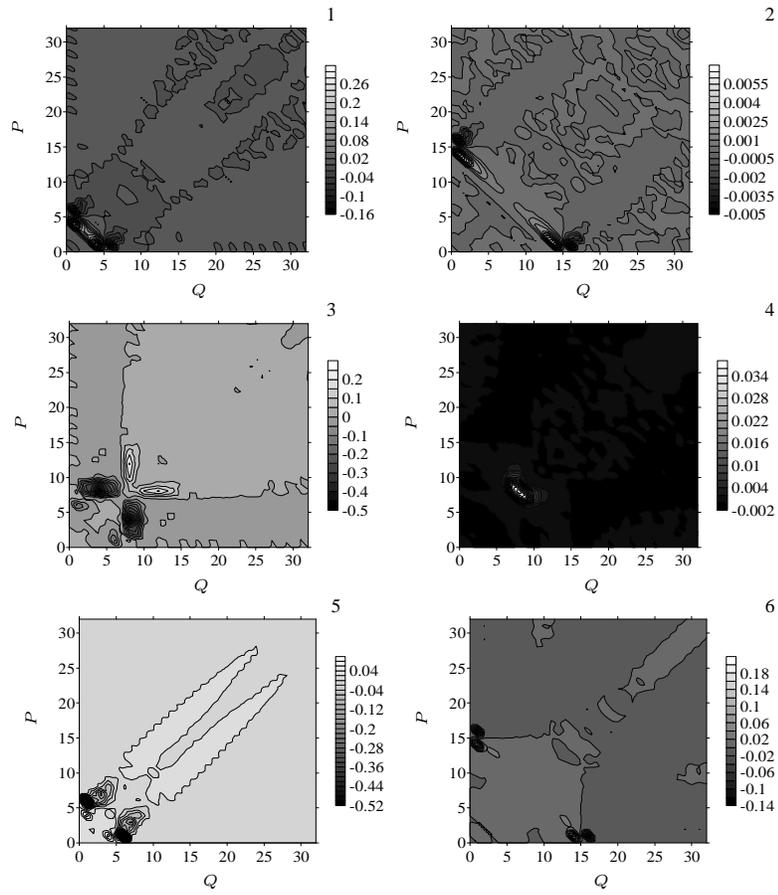,width=12cm}
\end{minipage}\hfill
\vskip 1.0cm
 \caption{ Fluxes  $T_3$ for fixed  $K$.
NR:   $K=5$ (graph 1), $K=15$ (2), R1:   $K=7$ (3), $K=15$ (4),
R2:  $K=5$ (5), $K=15$ (6) normalized at the maximal values
for all  $P$, $Q$ and $K$. } \label{fig8}
\end{figure}

Let us consider the properties of function  $T_3$ for three
regimes in more details.
 Because of the symmetry of the problem, we expect that  $T_3(K,P,Q)=T_3(K,Q,P)$,  which was
 used in constructing the discrete analog of the operator.   The case without rotation, see Fig.\ref{fig8}(1, 2) demonstrates a
 quite interesting result:   the largest input to  the energy flux for mode  $K$ comes from two sources:
 from the  $P\lesssim  K, Q\ll P$ and $Q\lesssim K, P\ll Q$.  In other words the wave triangle (K,P,Q) is  isosceles with a small angle between the equal sides  $\angle(K,P)$ or   $\angle(K,Q)$ .
 Taking into account that, according to Fig.\ref{fig7}, the  energy transfers to  $K$
 from the closest smaller wave number, we come to conclusion that the third small vector is a  catalyzer in the interaction while
participating in the interaction this high-frequency wave does
not provide to the wave $K$ itself with energy.
 This result is beyond the classical Kolmogorov scenario, according which the wave triangle is an
equilateral, i.e. not only the energy transfer is local, but the
interaction as well.  There are two reasons for this discrepancy.
Having in mind the results of \cite{ALEX07}, we connect this
result with a shortness of the considered spectra.
  On the other hand, it can be caused by anisotropy concerned  with the gravity force. In this connection,  it is known that, even
 for the larger Reynolds numbers in the problem with the imposed external force, the anisotropy on small scales can be substantial
 \citep{Zhou6}.

\begin{figure*}[t]
\vskip 1cm
\begin{minipage}[h!]{.45\linewidth}
\vspace*{2mm}
\begin{center}
\vskip -0cm \hskip -4cm
\includegraphics[width=9cm]{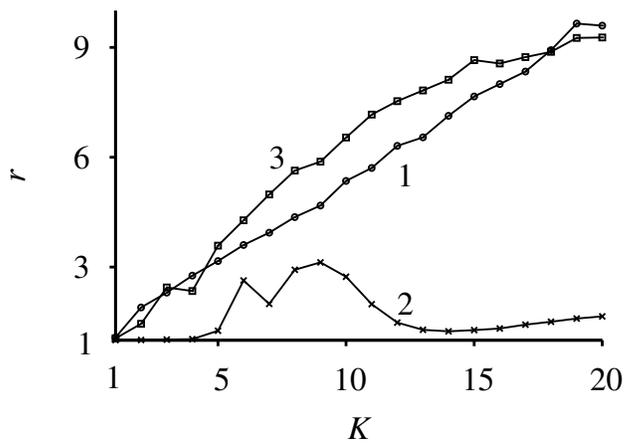}
\end{center}
\end{minipage}\hfill
\hskip -0cm
\begin{minipage}[h!]{.45\linewidth}
\vspace*{2mm}
\begin{center}
\vskip -0cm \hskip -2.7cm
\includegraphics[width=9cm]{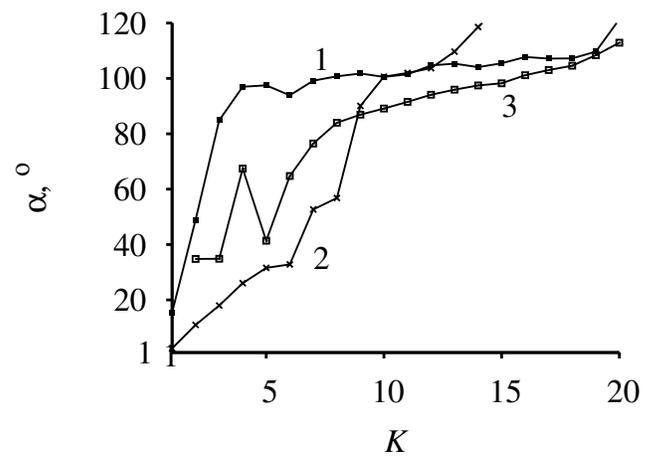}
\end{center}
\end{minipage}\hfill
\vskip 1cm \caption{Dependence maximum of the ratio  of  two wave
numbers  $P/Q,\, Q/P$ (left plot) and angle between
 vectors $\bf p$ and $\bf q$ (right plot), as a function of the resulting harmonic   $K$
 for regimes  NR (solid line), R1
(crosses) and R2 (diamonds).
 }
\label{fig9}
\end{figure*}

Our modeling demonstrates that there is a strong correlation of
the fluxes on small scales with a  buoyancy force on main scale.
 For this purpose, we introduce function   $\dsize r(K)=\int T_3\cdot  {\rm max}(P/Q, \, Q/P)\, d P\, dQ/ \int T_3\, d P\, dQ$ ($P,\, Q>0$), see Fig.\ref{fig9}.
 Virtually, for all $K$  the angle \\  $\dsize \alpha(K)= \int f\cdot T_3\, d P\, dQ /\int T_3\, d P\, dQ$ ($\dsize f=(P^2+Q^2-K^2)/2PQ$)
 between vectors  $\bf p$ and $\bf q$  is about $100^\circ$,  i.e. the modes, which take part in the interaction, are orthogonal in the wave space.

In the case with rotation the structure of the triangle with
small $\Ra$ differs from the case without rotation.
  The particular wave interacts with the set of the waves - see the cross-structure of the diagram in Fig.\ref{fig9}.
  This can be done by increasing the angle between vectors
 $\bf p$ and $\bf q$.  Note that $P\sim Q$ for small  $K$ and  $P$ , see  Fig.\ref{fig9}.

There is a well-pronounced inverse cascade of energy for small
$K$, see  Fig.\ref{fig8}:  harmonic  $K=7$
 receives energy from the higher wave numbers, transfering the energy, in its turn, to shorter waves.
  For  the larger $K$, there is a
   state with  $P\sim Q$ and  $\alpha\sim 100^\circ$.  In contrast to the case without rotation, the energy input to  $K$
  is made by waves  $P\sim Q\approx 0.7 K$.  The increase of  $\Ra$
 leads to the intermediate state.

 \section{Conclusion}\label{concl}
To summarize the main results of the paper:  the rotation
sufficiently changes the structure of the flows in the physical
space and its spectral properties. The introduction of rotation
leads to the transformation  of the typical Rayleigh-Benar
cellular convection to cyclonic form. In general, the rotation
suppresses convection due to enhanced dissipation: the horizontal
scale of the cyclone is $\E^{-1/3}$ times smaller than its
vertical scale
 $\rm L_z$. Moreover, the rotation leads to the selective suppression of the motion along the axis of rotation  ($\dsize {\lambda }_{\rm R2}\ll 1$)
 and  it also upsets the mirror-reflecting
 symmetry of the system, which is reflected in the non-zero mean hydrodynamic helicity  $\dsize\overline{ {\cal H}^{\cal H}}\ne 0$.

The behavior of the system in the regime with rotation with
moderate $\Ra$  is very different for $k>k_c$ and $k<k_c$.
 For   $k<k_c$ there is a weak inverse cascade with nonlocal interactions. It is possible to speak of statistical equilibrium when the energy
 exchange between the Fourier modes is absent. For  $k>k_c$, the cascade is direct, however, the nonlocal flux from small
 $k\sim k_c$ exists.    The slopes of the spectra for  $k< k_c$ and  $k> k_c$ are different.
 For long waves, the spectrum is close to white noise, and for large  $k$, the spectrum decays as $k^{-3}$.

We have demonstrated that, even in a pure hydrodynamic system
without a magnetic field, a variety of different interactions
occurs in the Fourier space. This analysis could be used for
adjusting semi-empirical turbulent models proposed  for geodynamo
purposes in the  future. In such models, there should be
agreement of the spectra and fluxes in the wave space for scales
larger than the cut-off scale $d_a$  with DNS of finer resolution.
The nontrivial point here is the reproduction of the inverse
cascades with  $d_c<d_a$.



\begin{singlespace}

\end{singlespace}

\newpage
\pagestyle{empty}

\newpage
\pagestyle{empty}

\newpage
\pagestyle{empty}

\eject
\newpage
\newpage
\pagestyle{empty}

\newpage
\pagestyle{empty}

\newpage
\pagestyle{empty}

\newpage
\pagestyle{empty}

\newpage
\pagestyle{empty}

\newpage
\pagestyle{empty}


\end{document}